# Comprehensive Review of Deep Reinforcement Learning Methods and Applications in Economics

Amir Mosavi [1,2,*], Pedram Ghamisi [3], Yaser Faghan [4], Puhong Duan [5]

1. Faculty of Civil Engineering, Technische Universität Dresden, 01069 Dresden, Germany;
2. Faculty of Humanities and Social Sciences, Oxford Brookes University, Oxford OX30BP, UK.
3. Exploration Devision, Helmholtz Institute Freiberg for Resource Technology, Helmholtz-Zentrum Dresden-Rossendorf, Dresden, Germany
4. Instituto Superior de Economia e Gestao, University of Lisbon, 1200-781, Lisbon, Portugal.
5. College of Electrical and Information Engineering, Hunan University, Changsha 410082, China; puhong_duan@hnu.edu.cn
* Correspondence: a.mosavi@brooks.ac.uk

**Abstract:** The popularity of deep reinforcement learning (DRL) methods in economics have been exponentially increased. DRL through a wide range of capabilities from reinforcement learning (RL) and deep learning (DL) for handling sophisticated dynamic business environments offers vast opportunities. DRL is characterized by scalability with the potential to be applied to high-dimensional problems in conjunction with noisy and nonlinear patterns of economic data. In this work, we first consider a brief review of DL, RL, and deep RL methods in diverse applications in economics providing an in-depth insight into the state of the art. Furthermore, the architecture of DRL applied to economic applications is investigated in order to highlight the complexity, robustness, accuracy, performance, computational tasks, risk constraints, and profitability. The survey results indicate that DRL can provide better performance and higher accuracy as compared to the traditional algorithms while facing real economic problems at the presence of risk parameters and the ever-increasing uncertainties.

**Introduction**

Deep learning (DL) techniques are based on the usage of multi-neuron that rely on the multi-layer architectures to accomplish a learning task. Where the neurons are linked to the input data in conjunction with loss function for the purpose of updating its weights and of maximizing the fitting to the inbound data [1, 2]. In the structure of multi-layer, every node takes the outputs of all the prior layers in order to represent outputs set by diminishing the approximation of the primary input data. While multiple-neurons perform to learn various weights on the same data at the same time. There is a great demand for the appropriate mechanisms to improve the productivity and product quality in the current market development. Deep learning enables us to predict and investigate complicated market trend compared to the traditional algorithms in ML. Where deep learning has the high potential to provide powerful tools to learn from stochastic data arising from multiple sources that can efficiently extract complicated relationships and features from given data to present better predictive tool to analyze the market [3, 4]. Additionally, compared to the traditional algorithms, deep learning is able to prevent over-fitting problem, to provide more efficient sample fitting associated with complicated interactions, and to outstretch input data to cover all the important feature of the relevant problem [5].

Reinforcement Learning [6] is a powerful mathematical framework for experience-driven autonomous learning [7]. Where the agents interact directly with the environment by taking actions to enhance its efficiency by trial-and-error with the purpose of optimizing the cumulative reward without requiring labelled data. Policy search and value function approximation are the key tools in reinforcement learning. While policy search is able to detect an optimal (stochastic) policy applying gradient-based or gradient-free approaches dealing with both continuous and discrete state-action settings [8]. The value function strategy is to estimate the expected return in order to find the optimal policy dealing with all possible actions based on the given state. While considering economic problem, despite traditional approaches [9], reinforcement learning methods prevent suboptimal performance, namely by imposing significant market constraints that lead to finding optimal strategy in terms of market analysis and forecast [10]. Despite RL successes in recent years [11-13], these results suffer the lack of scalability and disable to manage high dimensional problems. DRL technique by combining both RL and DL methods where DL equipped with the vigorous function approximation, representation learning properties of deep neural networks (DNN) and handling complex and nonlinear patterns of economic data, can efficiently overcomes these problems [14] [15]. Ultimately, the purpose of this paper is to comprehensively give the overview of the state-of-the-art in the application of both deep learning (DL) and deep reinforcement learning (deep RL) approaches in economics. However, in this paper, we focus on the state-of-the-art papers that employ DL, RL and DRL methods in economics issues. The main contributions of this paper can be summarized as follows:

- Classification of the existing DL, RL and DRL approaches in economics.
- Providing extensive insights into accuracy and applicability of DL, RL and DRL based economic models.
- Discussing the core technologies and architecture of DRL in the economic technologies.
- Proposing a general architecture of DRL in economic.
- Presenting open issues and challenges in current deep reinforcement learning models in economic.

The survey is organized as follows. We briefly review the common DL and DRL techniques in section 2 and 3, respectively. Section 4 proposes the core architecture and applicability of DL and DRL approaches in economic. Finally, we followed the discussion plus presenting the real-world challenges in DRL model in economics in section 5 and a conclusion to the work in section 6.

*Deep Learning Methods*

In this section, we review the most commonly used DL algorithms, which have been applied in various fields [16-20]. These deep networks comprise stacked auto-encoders (SAEs), deep belief networks (DBNs), convolutional neural networks (CNNs), recurrent neural networks (RNNs).

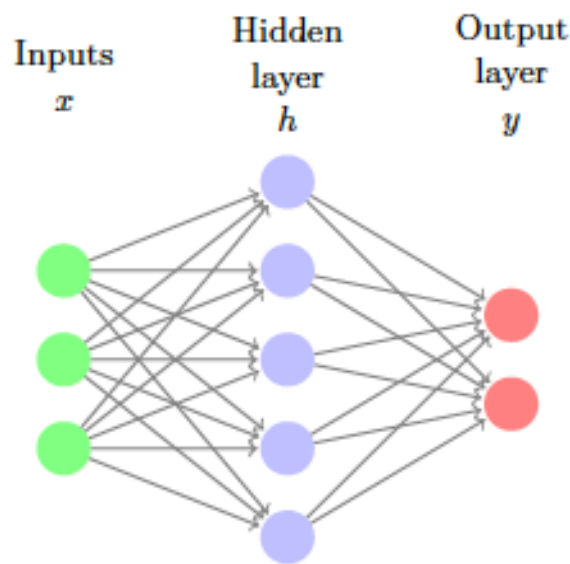

**Figure 1**. Structure of a simple neural network [20]

*Stacked Auto-Encoders (SAEs)*

The basic building block of the stacked AE is called auto-encoder (AE) which includes one visible inputs layer and one hidden layer [21]. It has two steps of training process. In mathematics, they can be explained as the Eq. 1 and 2:

$$h = f(w_h x + b_h) \qquad (1)$$

$$\mathcal{Y} = f(w_y x + b_y) \qquad (2)$$

The hidden layer $h$ can be transformed to provide the output values $\mathcal{Y}$. Here, $x \in \mathbb{R}^d$ is the input values, and $h \in \mathbb{R}^L$ denotes the hidden layer, i.e., encoder. $w_h$ and $w_y$ represent the input-to-hidden and hidden-to-input weights, respectively. $b_h$ and $b_y$ refer to the bias of the hidden and output terms, and $f(.)$ indicates an activation function. One can estimate the error term utilizing Euclidean distance for approximating input data $x$ while minimizing $\|x - y\|_2^2$.

*Deep Belief Networks (DBNs)*

The basic building block of a deep belief networks is known as restricted Boltzmann machine (RBM) which is a layer-wise training model [22]. It contains two-layer network with visible and hidden units. One can express the joint configuration energy of the units as Eq. 3:

$$E(v, h; \theta) = -\sum_{i=1}^{d} b_i v_i - \sum_{j=1}^{L} a_j h_j - \sum_{i=1}^{d}\sum_{j=1}^{L} w_{ij} v_i h_j = -b^T v - a^T h - v^T w h \qquad (3)$$

Where $b_i$ and $a_j$ are the bias term of the visible and hidden units, respectively. Here, $w_{ij}$ denotes the weight between the visible unit $i$ and hidden unit $j$. In RBM, the hidden units can capture an unbiased sample from the given data vector, as they are conditionally independent from knowing the visible states. One can improve the feature representation of a single RBM by cumulating diverse RBM one after another, which constructs a DBN for detecting a deep hierarchical representation of the training data.

*Convolutional Neural Networks (CNNs)*

CNNs are composed of a stack of periodic convolution layers and pooling layers with multiple fully connected layers. In the convolutional layer, CNNs employ a set of kernels to convolve the input data and intermediate features to yield various feature maps. In general, the pooling layer follows a convolutional layer, which is utilized to diminish the dimensions of feature maps and the network parameters. Finally, by utilizing the fully connected layers, these obtained maps can be transformed into feature vectors. We present the formula of the vital parts of the CNNs which are the convolution layers. Assume that $X$ be the input cube with the size of $m \times n \times d$ where $m \times n$ is regarding to spatial level of $X$, and $d$ count the channels. The $j$-th filter specified with the weight $w_j$ and bias $b_j$. Then, one can express the $j$-th output associated with the convolution layer as the Eq. 4:

$$y_j = \sum_{i=1}^{d} f(x_i * w_j + b_j), \ j = 1, 2, \dots, k \qquad (4)$$

Here the activation function $f(.)$ is used to enhance the network nonlinearity. Currently, ReLU [23] is the most popular activation function that leads to notably rapid convergence and robustness in terms of gradient vanishing [24].

*Recurrent Neural Networks (RNNs)*

RNNs extended the conventional neural network with loops in connections and were developed in [25], which can identify patterns in sequential data and dynamic temporal specification by utilizing recurrent hidden states compared to the feedforward neural network. Suppose that $x$ is the input vector. The recurrent hidden state $h^{<t>}$ of the RNN can be updated by Eq. 5:

$$h^{<t>} = \begin{cases} 0 & if\ t=0 \\ f_1(h^{<t-1>}, x^{<t>}) & otherwise \end{cases} \tag{5}$$

where $f_1$ denotes a nonlinear function, such as hyperbolic agent function. The update rule of the recurrent hidden state can be expressed as Eq. 6:

$$h^{<t>} = f_1(uh^{<t-1>} + wx^{<t>} + b_h), \tag{6}$$

where $w$ and $u$ indicate the coefficient matrices for the input at the current state and the activation of recurrent hidden units at the prior step, respectively. $b_h$ is the bias vector. Therefore, the output $y^{<t>}$ at time $t$ is presented as follows:

$$y^{<t>} = f_2(ph^{<t>} + b_y). \tag{7}$$

Here, $f$ is the nonlinear function, and $p$ is the coefficient matrix for the activation of recurrent hidden units at the current step and $b_h$ represents the bias vector. Due to the vanishing gradient of traditional RNN, long-term memory (LSTM) [26] and gated recurrent unit [27] were introduced to handle huge sequential data.

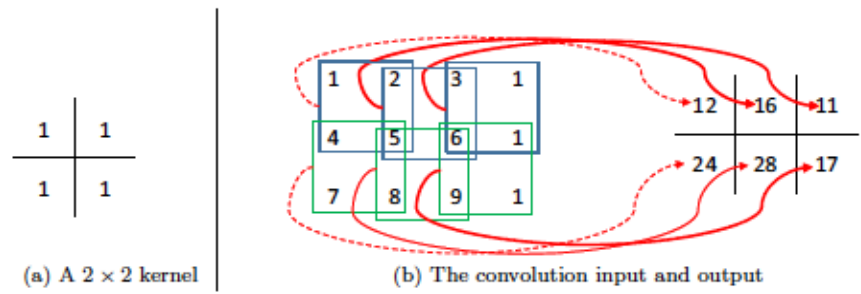

Figure 2. Image of convolution procedure [28]

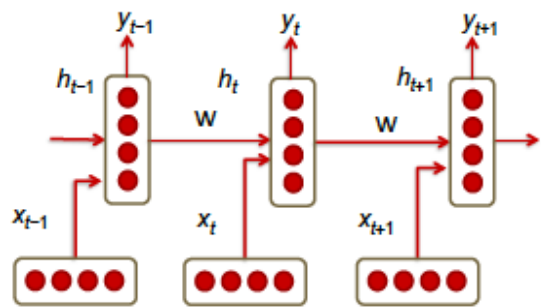

Figure 3. illustration of Recurrent Neural Network [29]

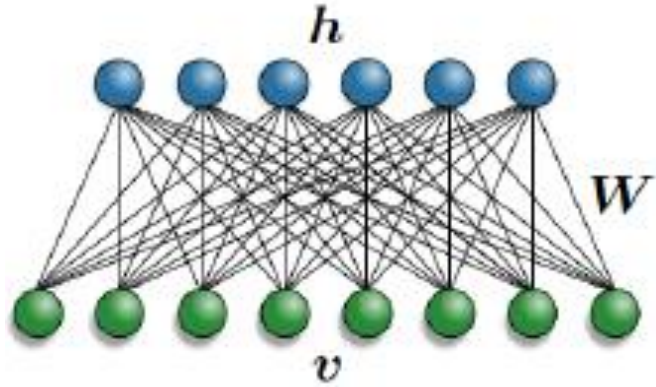

**Figure 4**. Image of RBM with the hidden layers [30]

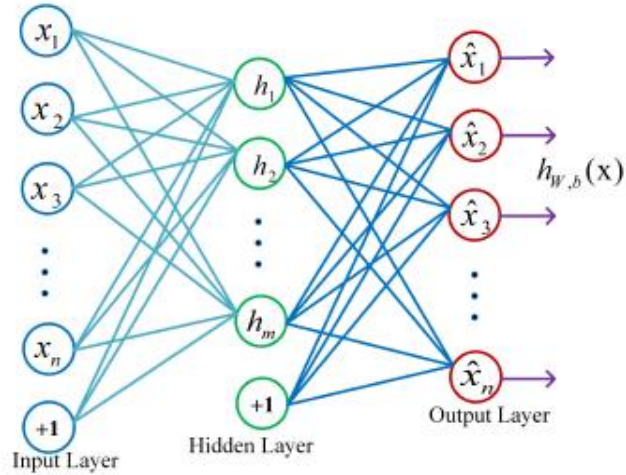

**Figure 5**. Simple architecture of Auto-Encoder [31]

*Deep Reinforcement Learning Methods*

In this section, we mainly focus on the most commonly used deep RL algorithms, such as value-based methods, policy gradient methods and model-based methods.

We first described how to formulate RL problem for the agent dealing with an environment while the goal is to maximize cumulative rewards. The two important characteristics of RL is, firstly that the agent has the capability of learning good behavior incrementally, secondly the RL agent enjoys the trial-and error experience with just dealing with the environment and gather information (see Figure 6). It is worth to mention that RL methods are able to practically provide most appropriate method in terms of computational efficiency as compared to some traditional approaches. One can model the RL problem with an Markov Decision Process (MDP) with a 5-tuple ($\mathcal{S}$, $\mathcal{A}$, T, $\mathcal{R}$, $\lambda$) where $\mathcal{S}$ (state space), $\mathcal{A}$ (action space), T $\in$ [0, 1] (transition function), $\mathcal{R}$ (reward function) and $\gamma \in [0, 1)$ (discount factor). The RL agent try to search a policy for the optimal expected return base on the value function $V^\pi(s)$ by Eq. 8:

$$V^\pi(s) = \mathbb{E}\left(\sum_{k=0}^{\infty} \gamma^k r_{k+t} \middle| s_t = s. \pi\right) \text{where } V^* = \max_{\pi \in \Pi} V^\pi(s) \tag{8}$$

Where:

$$r_t = \mathop{\mathbb{E}}_{a \sim \pi(s_t,.)} \mathcal{R}(s_t, a, s_{t+1}), \tag{9}$$

$$\mathbb{P}(s_{t+1}|s_t, a_t) = T(s_t. a_t. s_{t+1}) \text{ with } a_t \sim \pi(s_t. .) \tag{10}$$

Analogously, the Q value function can be expressed as the Eq. 11:

$$Q^\pi(s.a) = \mathbb{E}\left(\sum_{k=0}^{\infty} \gamma^k r_{k+t} \middle| s_t = s. a_t = a. \pi\right) \text{ where } Q^* = \max_{\pi \in \Pi} Q^\pi(s.a) \tag{11}$$

One can see the general architecture of the DRL algorithms in Figure 7.

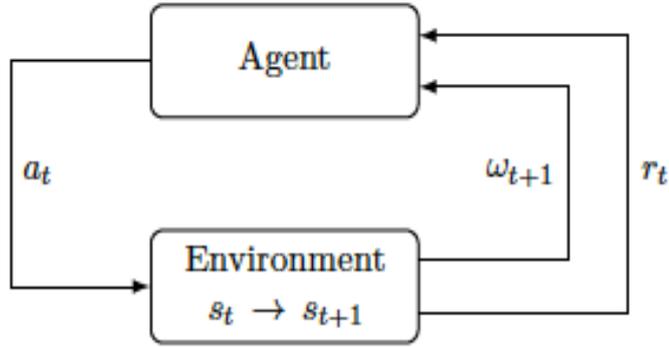

**Figure 6.** Interaction between agent and environment [20]

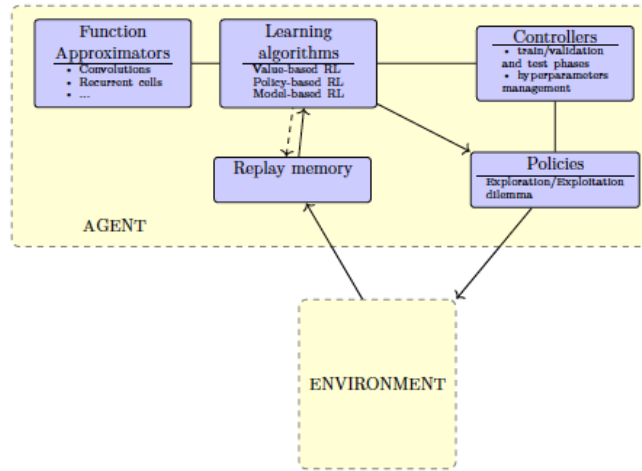

**Figure 7**. General structure of the deep reinforcement learning approaches [20]

Value-based Methods

The value-based class of algorithms allow us to construct a value function for defining a policy. We discuss the Q-learning algorithm [32], the deep Q-network (DQN) algorithm [6] with great success when playing ATARI games. We then give a brief review of the improved DQN algorithm.

Q-learning

The basic value-based algorithm called Q-learning algorithm. Assume Q be the value function (Equation 3.4), then the optimal value of the Q-learning algorithm using the Bellman equation [33] can be expressed as the Eq. 12:

$$Q^*(s.a) = (\mathcal{B}Q^*)(s.a) \tag{12}$$

Where the Bellman operator ($\mathcal{B}$) can be described as Eq.13:

$$(\mathcal{B}K)(s,a) = \sum_{\acute{s} \in S} T(s,a,\acute{s})(\mathcal{R}(s,a,\acute{s}) + \gamma \max_{\acute{a} \in \mathcal{A}} K(\acute{s},\acute{a})) \tag{13}$$

Here, the unique optimal solution of the Q value function is $Q^*(s, a)$. One can check out the theoretical analysis of optimality Q function in discrete space with sufficient exploration guarantee in [32]. In practice, a parameterized value function is able to overcome the high dimensional problems (possibly continuous space).

Deep Q-networks (DQN)

The DQN algorithm is presented by Mnih et al. [6] that can obtain good results for ATARI games in online framework. In deep Q-learning, we make use of a neural net to estimate a complex, nonlinear Q-value function. Imagine the target function as Eq. 14:

$$Y_k^Q = r + \gamma \max_{\acute{a} \in \mathcal{A}} Q(\acute{s}, \acute{a}; \bar{\theta}_k) \tag{14}$$

While $\bar{\theta}_k$ parameter, that define the values of the Q function at the $k^{th}$ iteration, will be updated just every $A \epsilon \mathbb{N}$ iteration to keep the stability and diminish the risk of divergence. In order to bound the instabilities, DQN uses two heuristics, i.e., the target Q-network and the replay memory [34]. Additionally, DQN takes the advantage of other heuristics like clipping the rewards for maintaining reasonable target values and for ensuring proper learning. An interesting aspect of DQN is that a variety of deep learning techniques are used to practically improve its performance such as preprocessing step of the inputs, convolutional layers and the optimization (stochastic gradient descent) [35]. We showed the general scheme of the DQN algorithm in Figure 8.

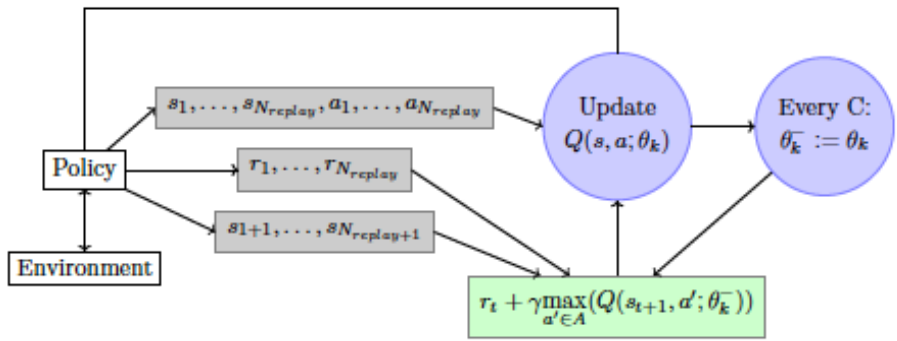

**Figure 8**. Basic structure of the DQN algorithm [20]

Double DQN

In Q-learning, the Q-value function utilizes the similar amount in order to pick out and to evaluate an action which one may cause overestimated values and upward bias in the algorithm respectively.

Thus, the double estimator method can be used for each variable to efficiently remove the positive bias in the action estimation process [36]. The Double DQN is independent of any source of error, namely stochastic environment error. The target value function in the Double DQN (DDQN) can be described as Eq. 15:

$$Y_k^{DDQN} = r + \gamma Q\left(\acute{s}, \arg\max_{a \in \mathcal{A}} Q\left(\acute{s}, a;\, \theta_k\right);\, \bar{\theta}_k\right) \tag{15}$$

Compared to Q-network, DDQN usually is able to improve stability and to obtain more accurate Q-value function as well.

Distributional DQN

The idea of approaches explained in the previous subsections were to estimate the expected cumulative return. Another interesting method is to represent a value distribution which allows us to better detect the inherent stochastic rewards and agent transitions in conjunction with the environment. One can define the random distribution return function associated with policy $\pi$ as follows Eq. 16:

$$Z^\pi(s, a) = \mathcal{R}(s, a, \acute{S}) + \gamma Z^\pi(\acute{S}, \acute{A}) \tag{16}$$

With random state-action pairs $(\acute{S}, \acute{A})$ and $\acute{A} \sim \pi(.\,|\acute{S})$. Thus, the Q value function can be expressed as the Eq. 17:

$$Q^\pi(s, a) = \mathbb{E}[Z^\pi(s, a)] \tag{17}$$

In practice, the distributional Bellman equation, interact with deep learning, can play the role of the approximation function [37-39]. The main benefits of this approach are implementation of the risk-aware behavior [40] and improved learning provides a richer set of training signals [41].

*Policy Gradient Methods*

This section discusses about policy gradient (PG) methods that are frequently used algorithms in reinforcement learning [42] which belongs to a class of policy-based methods. The method is to find a neural network parameterized policy in order to maximize the expected cumulative reward [43].

Stochastic Policy Gradient (SPG)

The easiest approach to obtain the policy gradient estimator could be to utilize algorithm [44]. The general approach to derive the estimated gradient is as Eq. 18:

$$\nabla_\omega \pi_\omega(s, a) = \pi_\omega(s, a)\, \nabla_\omega \log(\pi_\omega(s, a)) \tag{18}$$

While

$$\nabla_\omega V^{\pi_\omega}(s_0) = \mathbb{E}_{s \sim \rho^{\pi_\omega}, a \sim \pi_\omega}[\nabla_\omega (\log \pi_\omega(s, a))\, Q^{\pi_\omega}(s, a)] \tag{19}$$

Note that, in these methods the policy evaluation estimates Q-function and the policy improvement optimize the policy by taking a gradient step utilizing the value function approximation. The easy way of estimating the Q-function is to exchange it with a cumulative return from entire trajectories. A value-based method such as actor-critic methods can be used to estimate the return efficiently. In general, an entropy function can be used for the policy randomness and efficient exploration purpose. Additionally, it is common to employ an advantage value function where it does a measurement of comparison at the state to the expected return for each action. In practice, this replacement improves the numerical efficiency.

Deterministic Policy Gradient (DPG)

The DPG approach is the expected gradient of the action-value function. The deterministic policy gradient can be approximated without using an integral term over the action space. It can be demonstrated that the DPG algorithms can extremely perform better than SPG algorithms in high-dimensional action spaces [45]. NFQ and DQN algorithms can resolve the problematic discrete actions using the Deep Deterministic Policy Gradient (DDPG) [45] and the Neural Fitted Q Iteration with Continuous Actions (NFQCA) [46] algorithms, with the direct representation of a policy. An approach proposed by [47] to overcome the global optimization problem while updating greedy policy at each step. They defined a differentiable deterministic policy that can be moved to the gradient direction of value function for deriving the DDPG algorithm:

$$\nabla_\omega V^{\pi_\omega}(s_0) = \mathbb{E}_{s\sim\rho^{\pi_\omega}} [\nabla_\omega(\pi_\omega) \nabla_a(Q^{\pi_\omega}(s,a))| a = \pi_\omega(s)] \tag{20}$$

Which shows that the equation (2.25) is based on $\nabla_a(Q^{\pi_\omega}(s,a))$.

Actor-Critic Methods

An actor-critic architecture is a common approach where the actor updates the policy distribution with policy gradients and the critic estimate the value function for the current policy [48].

$$\nabla_\omega V^{\pi_\omega}(s_0) = \mathbb{E}_{s\sim\rho^{\pi_\beta}, a\sim\pi_\beta}[\nabla_\theta(\log \pi_\omega(s,a) Q^{\pi_\omega}(s,a))]. \tag{21}$$

where $\beta$ called behavior policy that makes the gradient biased and the critic with parameter $\theta$ estimate the value function, $Q(s,a;\theta)$, with the current policy $\pi$.

In deep reinforcement learning, the actor-critic functions can be parameterized with nonlinear neural networks [42]. Proposed approach by Sutton [7] was quite simple but not computationally efficient. The ideal is to design an architecture to profit from the reasonably fast reward propagation, the stability and the capability using replay memory. However, the new approach utilized in actor-critic framework presented by Wang et al. [49] and Gruslys et al. [50] have sample efficiency and computationally efficient as well.

Combining policy gradient and Q-learning

In order to improve the policy strategy in RL, an efficient technique needs to be engaged like policy gradient where applying a sample efficient approach and value function approximation associated with the policy. These algorithms enable us to work with continuous action spaces, to construct the policies for explicit exploration and to apply to multiagent setting where the problem deals with the stochastic optimal policy. However, the idea of combining policy gradient methods with optimal policy Q-learning was proposed by O Donoghue et al. [51] while summing the equation 2.24 with an entropy function (Eq.22):

$$\nabla_\omega V^{\pi_\omega}(s_0) = \mathbb{E}_{s,a}[\nabla_\omega(\log \pi_\omega(s,a)) Q^{\pi_\omega}(s,a)] + \alpha \mathbb{E}_s[\nabla_\omega H^{\pi_\omega}(s)] \tag{22}$$

Where
$$H^\pi(s) = -\sum_a \pi(s,a) \log \pi(s,a) \tag{23}$$

It showed that in some specific settings, both value-based and policy-based approaches have quite similar structure [52-54].

*Model-based Methods*

We have discussed so far, the value-based or the policy-based which belong to the method model-free approach. In this section, we focus on the model-based approach where the model deals with the dynamics of the environment and reward function.

Pure model-based methods

When the explicit model is unknown, it can be learned from experience by the function approximators [55-57]. The model plays the actual environment role to recommend an action. The common approach in the case of discrete actions is look ahead search, and trajectory optimization can be utilized in continuous case. A lookahead search is to generate potential trajectories with the difficulty of exploration and exploitation trade-off in sampling trajectories. The popular approaches to lookahead search are Monte Carlo tree search (MCTS) methods such that the MCTS algorithm recommends an action (see Figure 9). Recently, instead of using explicit tree search techniques, learning an end-to-end model was developed [58] with improved sample efficiency and performance as well. Lookahead search techniques is useless continuous environment. Another approach is PILCO so that apply Gaussian processes in order to learn a probabilistic model with reasonable sample efficiency [59]. However, the gaussian processes are just reliable in low-dimensional problem. One can take the benefit of the generalization capabilities of DL approaches to build the model of the environment in higher dimensions. For example, a DNN can be utilized in a latent state space [60]. Another approach aims at leveraging the trajectory optimization like guided policy search [61] by taking a few sequences of actions and then learns the policy from these sequences.

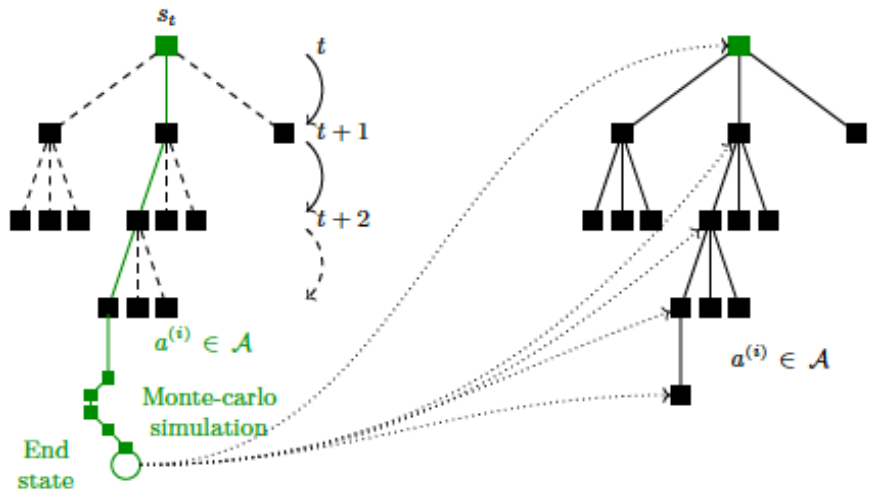

**Figure 9**. Illustration of the MCTS process [20]

Integrating model-free and model-based methods (IMF&MBM)

The situation to choose whether using the model-free versus model-based approaches mainly depends on the model architecture such as policy and value function. An example to clearly explain the key point. Assume that an agent needs to pass the street randomly while the best choice is to take the step unless something unusual happens in front of the agent. In this situation using model-based approach may be problematic due to the randomness model, while a model-free approach to find the optimal policy is highly recommended. There is a possibility of integrating planning and learning to achieve a practical algorithm which is computationally efficient. In the absence of a model where the limited number of trajectories is known, one approach is to build an algorithm to generalize well or to construct a model to generate more samples for model-free problem [62]. Other choice could be utilizing a model-based approach to accomplish primary tasks and apply model-free fine-tuning to achieve its goal successfully [63]. In the presence of a model, the tree search techniques directly [64]. The notion of neural network is able to combine these two approaches. The model proposed by Heess et al. [65] engaged backpropagation algorithm to estimate a value function. Another example is the work presented by Schema Networks [66] that uses prolific structured architecture where it leads to robust generalization.

**Review Section**

This section discusses an overview of various interesting utilization of both DL and deep RL approaches in economics.

*Deep Learning Application in Economics*

The recent attractive application of deep learning in variety of economics domains discussed in this section.

Deep Learning in Stock Pricing

In economical point of view, stock market value and its development are essential to business growth. In the current economic situation, there are many investors around the world that are interested in stock market in order to receive quick and better return compared to other sectors. In presence of uncertainty and risk in the forecasting of stock pricing, bring the challenges to the researcher to design market model for prediction. Despite all advances to develop mathematical models for forecasting, but still not that successful [67]. The deep learning topic attracts scientists and practitioner as it is useful to make huge revenue while enhancing the prediction accuracy with DL methods. Table 1 presents the summarized researches developed, recently.

Table 1. Application of deep learning in stock price prediction.

| References | Methods | Application |
|---|---|---|
| [68] | Two-streamed Gated recurrent unit network | Deep learning framework for Stock value prediction |
| [69] | Filtering methods | Novel filtering approach |
| [70] | Pattern techniques | Pattern matching algorithm for Forecasting stock value |
| [71] | Multilayer deep Approach | Advanced DL framework for stock value price |

According to Table 1, Minh et al [68] presented a more realistic framework for forecasting stock prices movement concerning financial news and sentiment dictionary as previous studies mostly relied on inefficient sentiment dataset, which are very crucial in stock trend, which lead to poor performance. They proposed Two-stream Gated Recurrent Unit (TGRU) by using deep learning techniques that performs better than LSTM model. Where it takes the advantage of applying two states that enable the model to provide much better information. They presented a sentiment Stock2Vec embedding with the proof of the model robustness in terms of market risk while using Harvard IV-4. Additionally, they provided a simulation system for the investors in order to calculate their actual return. Results were evaluated using accuracy, precision and recall values to compare TGRU and LSTM techniques with GRU. Fig. 10 presents the relative percentage of performance factors for TGRU and LSTM in comparison with those for GRU.

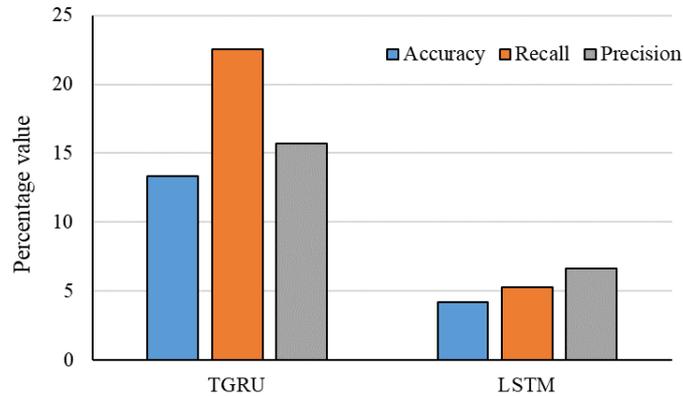

Fig. 10. Performance factors for comparing TGRU and LSTM with GRU

As is clear from Fig. 10, TGRU presents higher improvement in relative values for performance factor in comparison with LSTM. Also, TGRU provided higher improvement in recall values.
Song et al [69] presented a work to apply spotlighted deep learning techniques for forecasting stock trend. The research developed deep learning model with novel input-feature mainly focused on filtering technique in terms of delivering better training accuracy. Results were evaluated by accuracy values in training step. Comparing profit values for the developed approaches indicated that the novel filtering technology and stock price model by employing 715 features provided highest return value by more than 130$. Fig. 11 presents a visualized comparison for accuracy values.

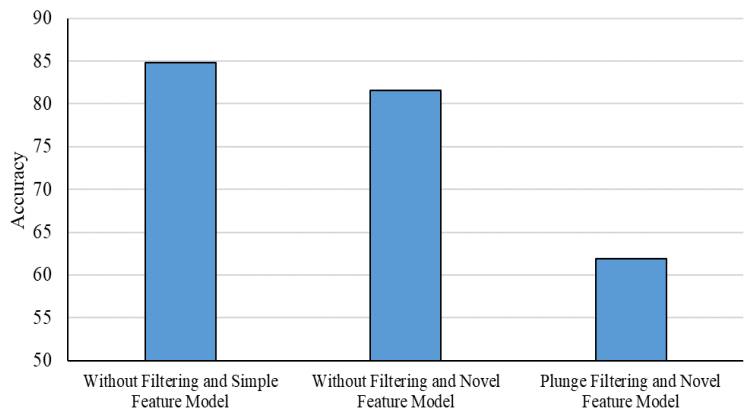

Fig. 11. Comparing accuracy values for the methods developed by Song et al [69]

Based on Fig. 11, the highest accuracy is related to simple feature model without filtering and the lowest accuracy is related to novel feature model with plunge filtering. By considering the trend, it can be concluded that, the absence of filtering process has a considerable effect on increasing the accuracy.
In another study, Go and Hong [70] employed DL technique to forecast stock value streams while analysing the pattern in stock price. The study, designed a DNN deep learning algorithm to find the pattern utilizing the time series technique which provided a high accuracy performance. Results were evaluated by percentage of test sets of 20 companies. Accuracy value for DNN was calculated to be 86%. But, DNN provided some disadvantages like over fitting and complexity. Therefore, it was proposed to employ CNN and RNN.
In the study by Das and Mishra [71] a new multilayer deep learning approach was used by employing time series concept for data representation to forecast close price of current stock. Results were evaluated by prediction error and accuracy values compared to the results obtained from the related studies. Based on results, the prediction error was very low based on outcome graph and as well as the predicted price was fairly close to reliable price in a time series data. Fig. 12 provides a comparison of the proposed method with the similar method reported by different studies in terms of accuracy.

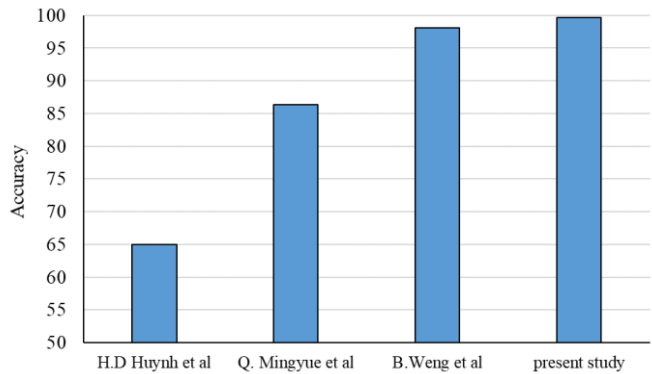

Fig. 12. The accuracy values for findings of the study by Das and Mishra [71]

Based on Fig.12, the proposed method by employing the related dataset could considerably improve the accuracy value by about 34, 13 and 1.5 % compared with Huynh et al [72], Mingyue et al [73] and Weng et al [74], respectively.

The used approach in [68] has the great advantage of utilizing forward and backward learning states at the same time to present more useful information. To explain briefly a part of mathematical forward pass formula for update gate is given (Eq.24):

$$\overrightarrow{z_t} = \sigma(\overrightarrow{W_z}x_t + \overrightarrow{U_z}\overrightarrow{h_{t-1}} + \overrightarrow{b_z}) \qquad (24)$$

And the backward pass formula is (Eq. 25):

$$\overleftarrow{z_t} = \sigma(\overleftarrow{W_z}x_t + \overleftarrow{U_z}\overleftarrow{h_{t-1}} + \overleftarrow{b_z}) \qquad (25)$$

where $x_t$ is the input vector, and b is the bias. $\sigma$ denotes the logistic function. $h_t$ is the activation function and W, U are the weights. In the construction of TGRU both forward and backward linked into a single context for the stock forecasting which lead to dramatically enhance the accuracy of the prediction. While applying more efficient financial indicators regarding to financial analysis. In Fig. 13 one can see the whole architectures of the proposed model.

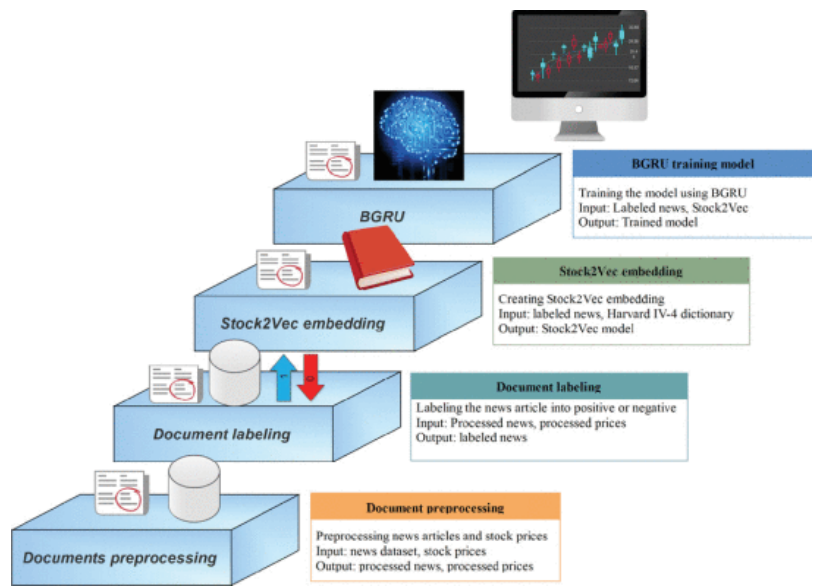

**Fig. 13**. Illustration of the TGRU structure [68]

Deep Learning in Insurance

Another application of DL methods is insurance sector. One of the attractive challenges that the insurance companies is to efficiently manage fraud detection (see Table 2). In recent years, ML techniques are widely used to develop practical algorithms in this field due to the high market demand for new approaches than traditional methods to practically measure all types of risks (Brockett et al. 2002; Pathak et al. 2005, Derrig, 2002). For instance, there are many demands for car insurance that forces the companies to deeply finding novel strategies in order to meliorate and upgrade their system. Table 2 summarizes the most notable studies for the application of DL techniques in insurance.

**Table 2.** Application of deep learning in Insurance industry.

| References | Methods | Application |
|---|---|---|
| [75] | Cycling algorithms | Fraud detection in car insurance |
| [76] | LDA-based appraoch | Insurance fraud |
| [77] | Autoencoder technique | Evaluation of risk in car insurance |

Bodaghi and Teimourpour [75] proposed a new method to detect professional frauds in car insurance for big data set by using social network analysis. Their approach employed cycling, which plays crucial roles in network system, to construct indirect collisions network and then identify doubtful cycles in order to make more profit concerning more realistic market assumption. Fraud detection may affect pricing strategies and long-term profit while dealing with insurance industry. Evaluation of the methods for suspicious components were performed by the probability of being fraudulent on the actual data. Fraud probability were calculated for different number of nodes in various community ID and cycle ID. Based on results, the highest Fraud probability for community was obtained at node number 10 by 3.759 and the lowest Fraud probability for community was obtained at node number 25 by 0.638. Also, the highest Fraud probability for cycle was obtained at node number 10 by 7.898 which was about 110 % higher than that for the community and the lowest Fraud probability for cycle was obtained at node number 12 by 1.638 which was about 156% higher than that for the community.

Recently, a new deep learning model presented to investigate fraud in car insurance bay Wang and Xu [76]. The proposed model outperform than traditional method where the combination of Latent Dirichlet Allocation (LDA) [66] and DNNs technique is utilized to extract the text features of the accidents comprising traditional features and text features. Another important topic that brings interest to insurance industry is telematics devices which deal with detecting hidden information inside the data efficiently. Results were evaluated by Accuracy and precision performance factors in two scenarios including "with LDA" and "without LDA" to consider the effect of LDA on prediction process. Fig. 14 presents the visualized results.

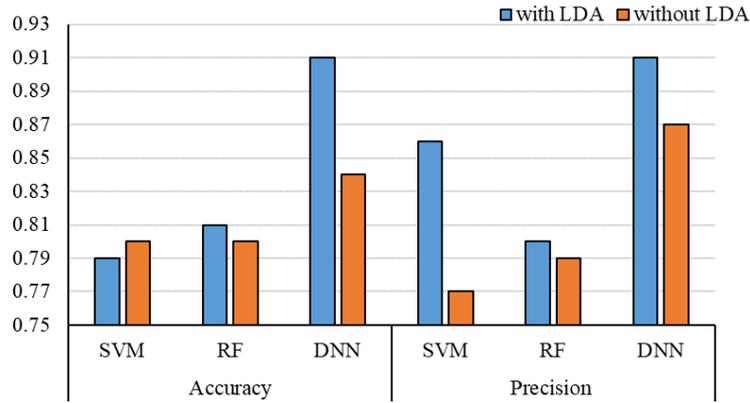

Fig. 14. The results of the study by Wang and Xu [76]

According to Fig. 14, LDA has a positive effect on accuracy of DNN and RF that could successfully increase the accuracy of DNN and RF by about 7% and 1%, respectively but reduced the accuracy of SVM by about 1.2%. In the other hand, LDA could successfully increase the precision of SVM, RF and DNN by about 10, 1.2 and 4 %, respectively.

A recent work proposed an algorithm combining auto-encoder technique with telematics data value to forecast the risk associated with insurance customers [77]. To efficiently dealing with large dataset one requires powerful updated tool for detecting valuable information from it like telematics devices [77]. They utilized conceptual model in conjunction with telematics technology to forecast the risk (see Fig. 15). While the risk score (RS) compute as Eq. 26:

$$(RS)_j = \sum_i W_{c_i} * O_{ij} \quad \text{where} \quad W_{c_i} = \frac{\sum_i \sum_j O_{ij}}{\sum_j O_{ij}} \tag{26}$$

Where $W$ and $O$ present the risk weight and driving style, respectively. Their experimental results showed the superiority of their proposed approach.

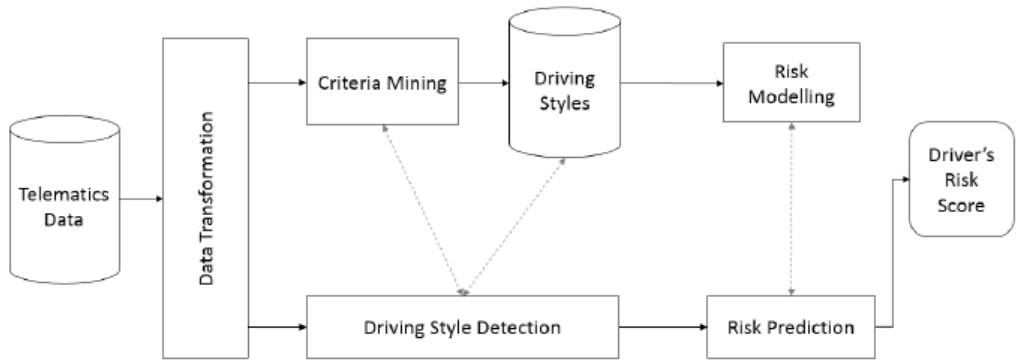

**Fig. 15**. The graph of the conceptual model [77]

Deep Learning in Auction Mechanisms

Auction design has a major importance in practice that allows the organizations to present better services to their customers. A great challenge to learn a trustable auction is that its bidders require to project optimal strategy for maximizing the profit. In this direction, Myerson designed an optimal auction with only single item [78]. There are many works with results for single bidders but most often with partial optimality [79-81]. Table 3 presents the notable studies developed by DL techniques in Auction Mechanisms.

Table 3. Application of deep learning in auction design.

| References | Methods | Application |
|---|---|---|
| [82] | Augmented Lagrangian Technique | Optimal auction design |
| [83] | Extended RegretNet method | Maximized return in auction |
| [84] | Data-driven Method | Mechanism design in auction |
| [85] | Multi-layer neural Network method | Auction in mobile network |

Dütting et al [86] designed a compatible auction with multi-bidder that maximize the profit by applying multi-layer neural networks for encoding its mechanisms. The proposed method was able to solve much more complex tasks while using augmented Lagrangian technique than LP-based approach. Results were evaluated by comparing the total revenue. Despite all previous results, the proposed approach provided a great capability to be applied to the large setting with high profit and low regret. Fig. 16 presents the revenue outcomes for the study by Dütting et al [86].

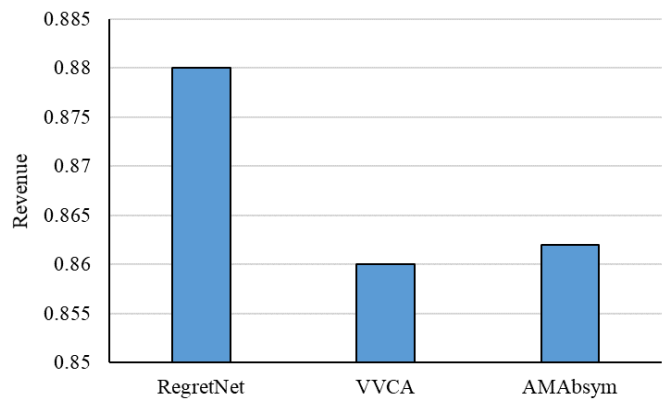

Fig. 16. The comparison of revenue for the study by Dütting et al [86]

According to Fig. 16, RegretNet as the proposed technique increased the revenue by about 2.32 and 2 % compared with VVCA and AMAbsym, respectively.

Another work employed deep learning approach to extend the result in [82] in terms of both budget constraints and Bayesian compatibility [83]. Their method demonstrated that neural networks are able to efficiently design novel optimal-revenue auctions by focusing on multiple setting problems with different valuation distributions. Additionally, a new method proposed by [84] improves the result in [86] by constructing a different mechanisms where it applied deep learning techniques. Their approach makes use of strategy-profess under assumption that multiple bids can be applied to each bidder. Other attractive approach applied to mobile blockchain networks, constructed an effective auction using multi-layer neural network technique [85]. While neural networks trained by formulating parameters in order to maximize the profit of the problem, which was shown that considerably outperformed than the baseline approach. The main recent work mentioned in Table 3 indicate that this field is growing fast.

The proposed approach in [83] modified the regret definition in order to namely handle budget constraints while designing an auction with multiple items settings (see Figure 12). The main expected regret is as Eq. 27:

$$rgt_i = \mathbb{E}_{t_i \sim F_i}\left[\max_{t'_i \in \mathcal{T}_i} \chi\left(p_{i(t'_i)} \leq b_i\right)\left(\mathcal{U}_i(t_i, t'_i) - \mathcal{U}_i(t_i, t_i)\right)\right], \quad (27)$$

Where $\chi$ is the indicator function, $\mathcal{U}$ is the interim utility and $p$ is the interim payment. Their method improved the state-of-art utilized DL concepts to optimally design an auction applied to the multiple items with high profit.

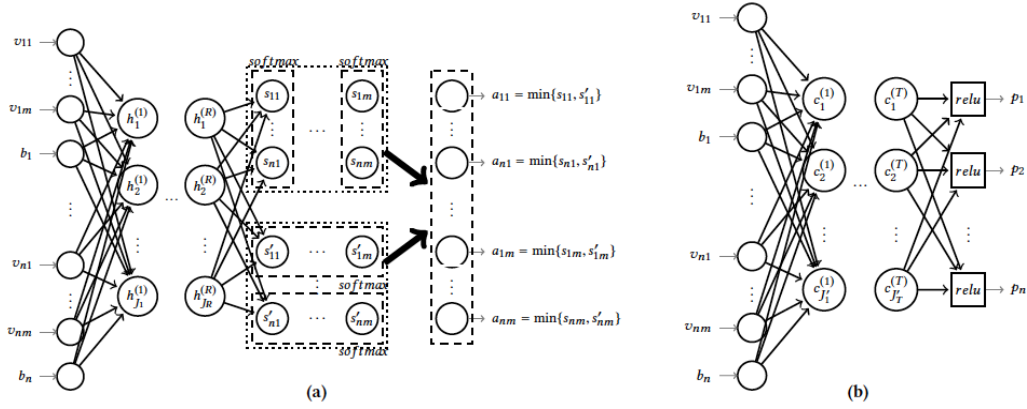

**Figure 12**. Illustration of budgeted RegretNet [83]

Deep Learning in Banking and Online Market

In current technology improvement, fraud detection is one the challenging application of deep learning, namely in online shopping and in credit card. There is a high market demand to construct an efficient system for fraud detection in order to keep the involved system safe (see Table 4).

Table 4. Application of deep learning in banking system and online market.

| References | Methods | Application |
| --- | --- | --- |
| [87] | AE | Fraud detection in unbalanced datasets |
| [88] | Network topology | credit card transactions |
| [89] | Natural language Processing | Anti-money laundering detection |
| [90] | AE and RBM architecture | Fraud detection in credit card |

An unsupervised learning could be used to investigate online transactions due to variable patterns of frauds and change of customer's behavior. An interesting work relied on employing deep learning methods such as AE and RBM to mimic irregularity from regular pattern by rebuilding regular transactions in real-time [90]. They applied fundamental experiments to confirm that AE and RBM approaches are able to accurately detect credit card with the huge dataset. Although, deep learning approaches enable us to fairly detect the fraud problem in credit card, but to build the model make use of diverse parameters that affect its outcomes. The work of Abhimanyu [84] made the evaluation of common used methods in deep learning to efficiently check out previous fraud detection problems in terms of class inconsistency and scalability. A plenary advice is provided by the authors regarding to analyzing model parameters sensitivity and its tuning while applying neural network architecture to the fraud detection problems in credit card. Another research proposed by [87] designed an autoencoder algorithm in order to model the fraudulent activities as efficient automated tools need to accurately handle huge daily transactions around the world. The model enables investigators to give report regarding to unbalanced datasets where there is no need of using data balance approaches like Under-Sampling approach. One of the world largest industry called money laundering, which is the unlawful process of hiding the original source of received money unlawfully by transmitting it through a complicated banking transaction. A recent work, considering anti money laundering detection, designed a new framework using natural language processing (NLP) technology [89]. Here, the main reason of constructing deep learning framework is due to diminishing the cost of human capital and consuming time. Their distributed and scalable method (e.g. NLP) preforms complex mechanism associated with various data source like news and tweets in order to make the decision simpler while providing more record.

It is a great challenge to design practical algorithm to prevent fraudulent transaction in financial sectors while dealing with credit card. The work in [87] presented efficient algorithm that has the superiority of controlling unbalanced data compared to traditional algorithms. Where anomalies detection can be handled by the reconstruction loss function. We depicted their proposed AE algorithm in Table 5.

**Table 5.** AE pseudo algorithm [87].

| Steps | Processes |
|---|---|
| Step 1: Prepare the input date | Input Matrix X II input dataset<br>Parameter of the matrix//parameter (w,bx,bh)<br>where: w : Weight between layers, bx Encoder's parameters, $b_h$ Decoder's Parameters |
| Step 2: initial Variables | h←null // vector for hidden layer<br>X←null // Reconstructed x<br>L←null II vector for Loss Function<br>1←batch number<br>i←0 |
| Step 3: loop statement | While i< 1 do<br>II Encoder function maps an input X to hidden representation h:<br>h = f (p[i ].w+ p[i] bx)<br>/* Decoder function maps hidden representation h back to a Reconstruction X :*/<br>X = g(p[i ].l / +p[ i]bx)<br>/*For nonlinear reconstruction, the reconstruction loss is generally from cross-entropy :*/<br>L = -sum(x *log(X) + (1-x ) *log(l -X ))<br>/* For linear reconstruction, the reconstruction loss is generally from the squared error:*/<br>L = sum(x -X)$^2$<br>Min<br>Θ[i]= p L(x -X)<br>End while<br>Return Θ |
| Step 4: output | Θ←<null matrix>//objective function<br>/*Training an auto-encoder involves finding parameters = (W,hx , bb) that minimize the reconstruction loss on the given dataset X and the objective function*/ |

Deep Learning in Macroeconomic

Macroeconomic prediction approaches have got much interest in recent years which are helpful for investigating economics growth and business changes [91]. There are many proposed methods that can forecast macroeconomic indicators but these approaches require huge amounts of data and suffer from model dependency. One can check out in Table 6 to see the recent results which are more acceptable than the previous ones.

Table 6. Application of deep learning in macroeconomic.

| References | Methods | Application |
|---|---|---|
| [92] | Encoder-decoder | Indicator prediction |
| [93] | Backpropagation Approach | Forecasting inflation |
| [94] | Feed-forward neural Network | Asset allocation |

Smalter and Cook [92] presented a new robust model called encoder-decoder that make use of deep neural architectures to prosper the accuracy of prediction with low data demand concerning unemployment problem. Mean absolute error (MAE) value was employed for evaluating the results. Fig. 13 visualizes the average values of MAE. This visualization compares average MAE values for CNN, LSTM, AE, DARM and Survey of Professional Forecasters (SPF).

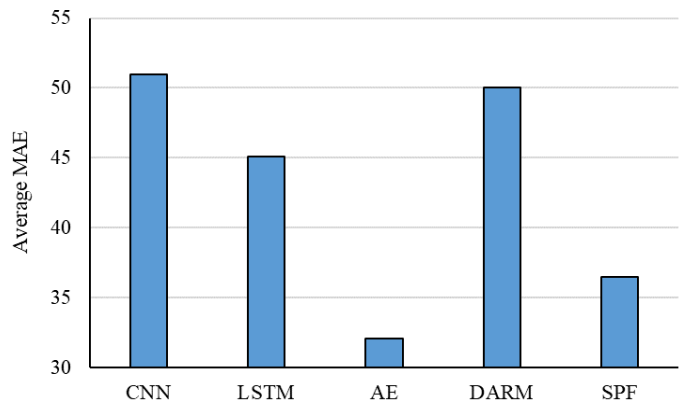

Fig. 13. Average MAE reports for the study by Smalter and Cook [88]

According to Fig. 13, the lowest average MAE is related to AE followed by SPF with additional advantages such as supplying nice single-series efficiency, higher accuracy of predicting and better model specification.

Haider and Hanif [93] employed an ANN method that significantly applied to forecast macroeconomic indicators for inflation. Results were evaluated by RMSE values for comparing the performance of ANN, AR and ARIMA techniques. Fig. 14 presents the average RMSE values for comparison step.

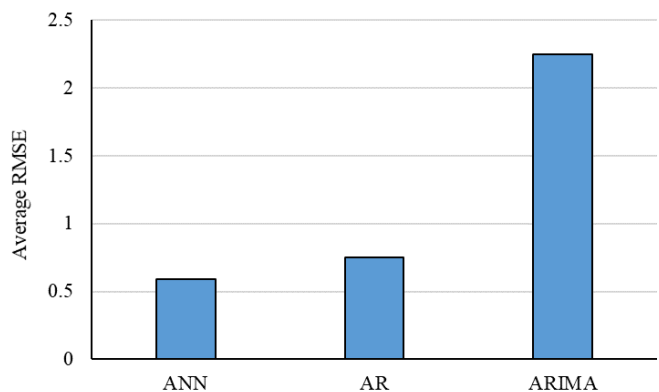

Fig. 14. Average RMSE reports for the study by Haider and Hanif [93]

According to Fig. 14, the study used model simulation based on backpropagation approach outperforms than previous models like autoregressive integrated moving average (ARIMA) model. Another useful application of DL architecture is to deal with investment decisions in conjunction with macroeconomic data. Chakravorty et al [94] used feed-forward neural network to perform tactical asset allocation while applying macroeconomic indicators and price-volume trend. They proposed two different methods in order to build a portfolio while the first one applied to estimate expected returns and uncertainty and the second approach used to obtain allocation directly using neural network architecture and optimize portfolio Sharpe. Their methods with the adopted trading strategy, demonstrated that have a comparable achievement with previous results.

A new technique is used in [92] to enhance the indicator prediction accuracy while the model requires low data. Experimental results indicate that encoder-decoder totally outperform the highly cited SPF prediction. The results in Table 7 showed that the encoder-decoder is more responsive than the SPF prediction or is more adaptable to the data changes.

**Table 7**. Inflection points prediction for Unemployment near 2007 [92]

| Time horizon | SPF | Encoder-Decoder |
|---|---|---|
| 3month horizon model | Q3 2007 | Q1 2007 |
| 6month horizon model | Q3 2007 | Q2 2007 |
| 9month horizon model | Q2 2007 | Q3 2007 |
| 12month horizon model | Q3 2008 | Q1 2008 |

Deep Learning in Financial Market (Service & risk management)

In financial market, it is crucial to efficiently handle the risk arising from credits. Due to recent advance in big data technology, DL models can design a reliable financial models in order to forecast credit risk in banking system (see Table 8).

**Table 8.** Application of deep learning in financial market (services and risk management).

| References | Methods | Application |
|---|---|---|
| [95] | Binary Classification Technique | Loan pricing |
| [96] | Feature selection | Credit risk analysis |
| [97] | AE | Portfolio management |
| [98] | Likelihood Esrtimation | Mortgage risk |

Addo et al [95] employed binary classification technique to give essential features of selected ML and DL models in order to evaluate the stability of the classifiers based on their performance. The study used the models separately to forecast loan default probability, by considering the selected features and the selected algorithm, which are the crucial keys in loan pricing processes. In credit risk management it is important to distinguish the good and the bad customers which oblige us to construct very efficient classification models by using deep learning tool. Fig. 15 presents the results visualized for the study by Addo et al [95] to compare the performance of LR, RF, Gboosting and DNN in terms of RMSE and area under the curve values.

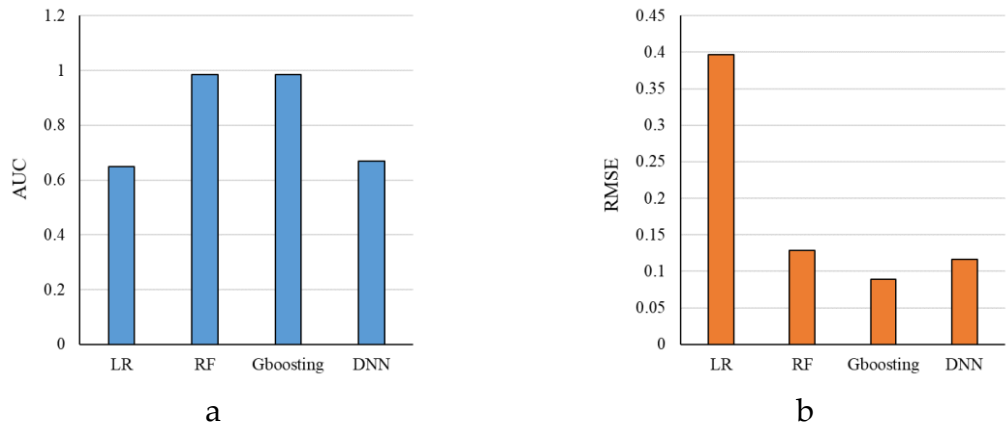

Fig. 15. AUC (a) and RMSE (b) results for the study by Addo et al [95]

In general, credit data include useless and unneeded features that require to be filtered by the feature selection strategy to provide the classifier with high accuracy.

Ha and Nguyen [96] studied novel feature selection method that help financial institution to perform credit assessment with less workload while focusing on important variables and to improve the classification accuracy in terms credit scoring and customer rating. Accuracy value was employed for comparing the performance of Linear SVM, CART, k-NN, Naïve Bayes, MLP and RF techniques. Fig. 16 presents the comparison results related to the study by Ha and Nguyen [96].

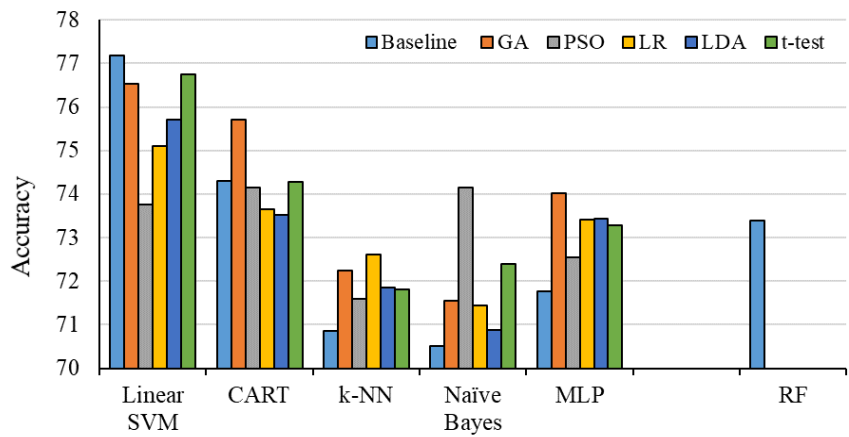

Fig. 16. The comparison results for the study by Ha and Nguyen [96]

Fig. 16 contains the base type of methods and their integrating with GA, PSO, LR, LDA and t-test as different scenarios except RF which is in its baseline type. In all the cases except linear SVM, integrating with optimizers and filters like GA, PSO and LR improved the accuracy value compared with their baseline type. But, in SVM, the baseline model provided the higher accuracy compared with other scenarios.

A research applied hierarchical models to present high performance regarding to big data [97]. They construct a deep protfolio with the four process of auto-encoding, calibrating, validating and verifying to the application of a portfolio including put and call options with underlying stocks. Their methods are able to find optimal strategy in the scene of the theory of deep portfolios by ameliorating the deep feature. Another work developed deep learning model in mortgage risk applied to huge data sets that discovered the nonlinearity relationship between the variables and debtor behavior which can be affected by the local economic situation [98]. Their research demonstrated that the unemployment variable is substantial in the risk of mortgage which brings the importance of implications to the relevant practitioners.

Deep Learning in Investment

Financial problems generally require to be analyzed in terms of dataset from multiple sources. Thus, it is substantial to construct a reliable model for handling unusual interaction and features from the data for efficient forecasting. Table 10 comprises the recent result of using deep learning approaches in financial investment.

Table 10. Application of deep learning in stock price prediction.

| References | Methods | Application |
|---|---|---|
| [99] | LSTM and AE | Market investment |
| [100] | Hyper-parameter | Option pricing in finance |
| [101] | LSTM and SVR | Quantitative strategy in investment |
| [102] | R-NN and genetic Method | Smart financial investment |

Aggarwal and Aggarwal [99] designed a deep learning model that applied to economical investment problem with the capability of extracting nonlinear data pattern. They presented a decision model using neural network architecture such as LSTM, auto-encoding and smart indexing to better estimate the risk of portfolio selection with securities for the investment problem. Culkin and Das [100] devoted to option pricing problem using DNN architecture to reconstruct well-known Black and Scholes formula with considerable accuracy. The study tried to revisit the previous result [103] and made the model more accurate with the diverse selection of hyper-parameters. The experimental result displayed that the model can price the options with minor error. Fang et al [101] investigated the option pricing problem in conjunction with transaction complexity where the research goal is to explore efficient investment strategy in high frequency trading manner. The work used delta hedging concept to handle the risk in addition to the several key components of the pricing such as implied volatility and historical volatility, etc. LSTM-SVR models applied to forecast the final transaction price with experimental results. Results were evaluated by deviation value and were compared with single RF and single LSTM methods. Fig. 17 presents the visualized results to compare results more clear for the study by Culkin and Das [100].

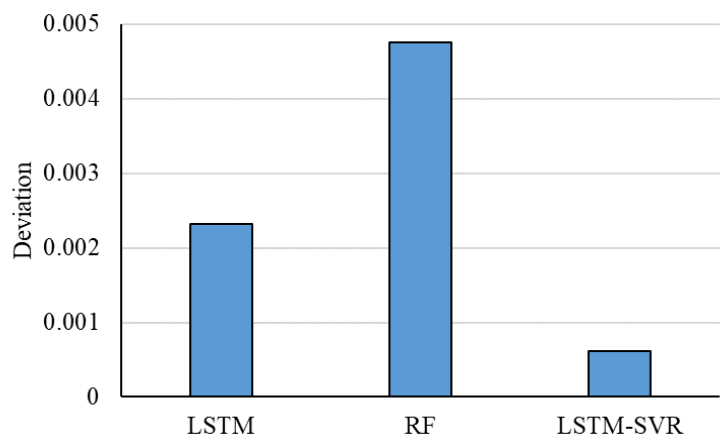

Fig. 17. Deviation results for the study by Culkin and Das [100]

Based on results, the single DL approach outperforms the traditional RF approach with higher return in adopted investment strategy, but its deviation was considerably higher than that for the hybrid DL (LSTM-SVR) method. A novel learning Genetic Algorithm is proposed by Serrano [102] concerning Smart Investment that use R-NN model to emulate the human behavior. The model makes usage of complicated deep learning architecture where reinforcement learning installs for the fast decisions-making purpose, deep learning for constructing stock identity, clusters for the overall decisions-making purpose and genetic for transferring purpose. Their created algorithm improved the return regarding to the market risk with minor error performance.

The authors in [102] construct complex model while genetic algorithm using network weights to imitate the human brain where applying the following formula for the encoding-decoding organism (Eq. 28):

$$min\|X - \zeta(W_2\zeta(XW_1))\| \quad \text{s.t.} \quad W_1 \geq 0 \tag{28}$$

and

$$W_2 = \text{pinv}(\zeta(XW_1))\,X \quad, \quad \text{pinv}(x) = (x^T x)x^T \tag{29}$$

Where $x$ is the input vector, $W$ is the weights matrix. $Q_1 = \zeta(XW_1)$ represents the neuron vector and $Q_2 = \zeta(W_2 Q_1)$ represents the cell vector. We illustrate the whole structure of the model in Fig. 18.

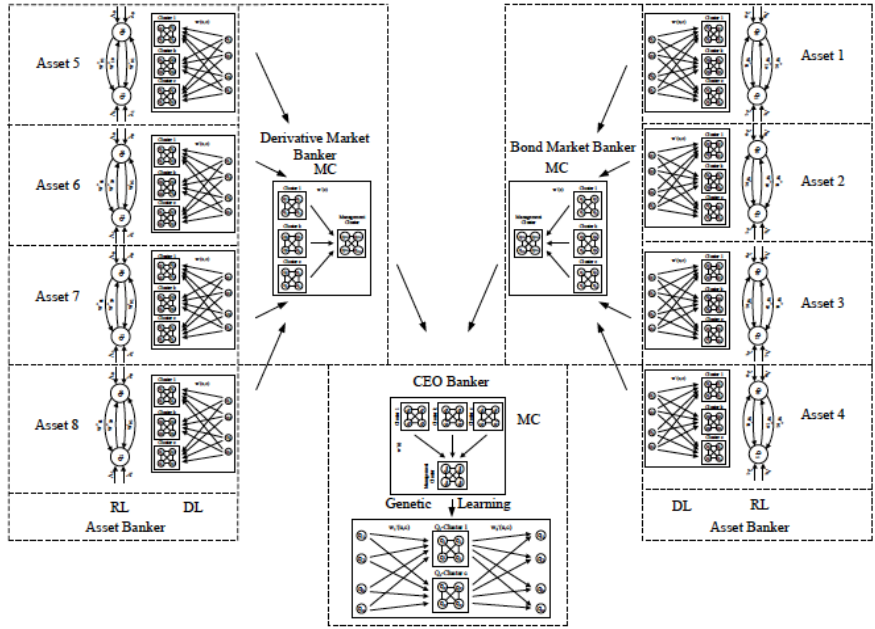

**Fig. 18**. Structure of smart investment model [102]

Deep Learning in Retail

Augmented reality (AR) enables the customers to improve their experience while buying/finding a product from the real stores. This algorithm was frequently used by researchers in the field. Table 11 presents the notable studies.

**Table 11.** Application of deep learning in retail market.

| References | Methods | Application |
|---|---|---|
| [104] | Augmented reality And image Classification | Improving shopping in retail market |
| [105] | DNN methods | Sale prediction |
| [106] | CNN | Investigation in retail stores |
| [107] | Adaptable CNN | Validation in food industry |

Cruz et al [104] in a study combined DL technique and augmented reality approaches in order to cater the fruitful information for the clients. They presented a mobile application such that it is able to locate the clients by means of the image classification technique in deep learning. Then, by using new AR approaches to efficiently advise finding the selected product with all helpful information regarding to that product in large stores. Another interesting application of deep learning methods is in fashion retail with large supply and demand where precise sales prediction is tied to the company's profit. Nogueira et al [105] designed a novel DNN to accurately forecast future sales where the model uses a huge and completely different set of variables such as physical specifications of the product and the idea of subject-matter expert. Their experimental work indicated that DNN model performance is comparable with other shallow approaches like RF, SVR. An important issue for the retail persons is to analyze client behavior in order to maximize their revenue that can perform by the computer vision techniques in the study by De Sousa Ribeiro et al [106]. The proposed CNN regression model to deal with counting problem where assessing the number of available persons in the stores and detecting crucial spots. The work also designed foreground/background approach for detecting the behavior of the persons in retail markets. Results of the proposed method was compared with a common CNN technique in term of accuracy. Fig. 18 presents the visualized result for the study by De Sousa Ribeiro et al [106].

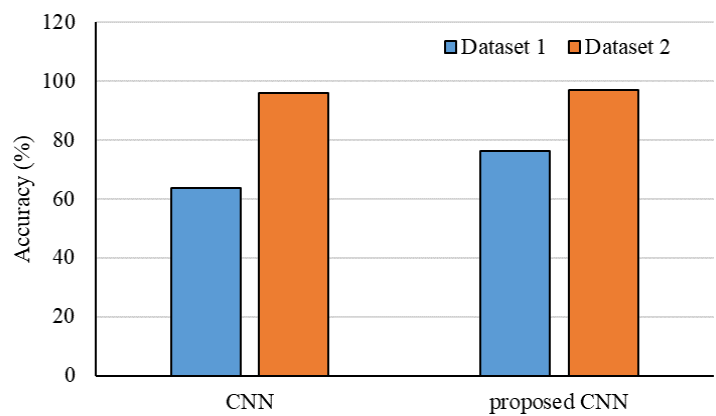

Fig. 18. Results for the study by De Sousa Ribeiro et al [106]

According to Fig. 18, the proposed model is robust enough and perform better than common CNN methods in both datasets. In the current situation of the food retail industry, it is crucial to distribute the products to the market with proper information and remove the possibility of mislabeling in terms of public health.

A new adaptable convolutional neural network approach for Optical Character Verification, presented with aim to automatically identify the use by the dates in large dataset [107]. The model applies both a k-means algorithm and k-nearest neighbor to incorporate calculated centroids to the both CNN for the efficient separation and adaptation. Developed models allow us to better manage the precision and the readability of important use related to dates and as well as useful to better control the tractability information in food producing processes. The most interesting recent research collected in Table 11 with the application of DL in retail market.

The adaptation method in [117] interconnects C and Z cluster centroids illustrated in Figure 11. This is to construct new augmented cluster including information from both networks in CNN1 and CNN2 using k-NN classification (see Fig. 19). Their methodologies lead to considerably improve food safety legally and verification correct labelling.

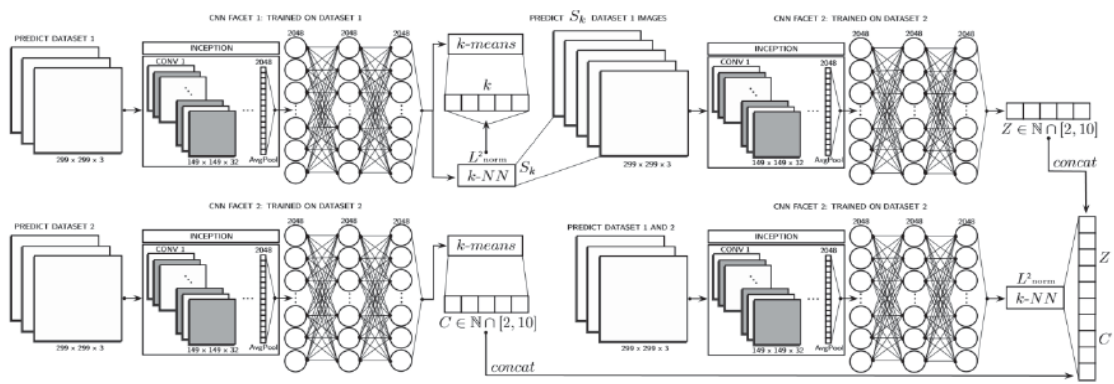

**Fig. 19**. Full structure of the adaptable CNN framework [107]

Deep Learning in Business (Intelligence)

Nowadays, big data solutions play the key role in the business services and productivities to efficiently reinforce the market. To solve the complex business intelligence (BI) problems dealing with market data, DL techniques is useful (see Table 12). Table 12 presents the notable studies for the application of deep learning in business intelligence.

Table 12. Application of deep learning in business intelligence.

| References | Methods | Application |
| --- | --- | --- |
| [107] | MLP | BI with client data |
| [31] | MLS and SAE | Feature selection in market data |
| [109] | RNN | Information detection in business data |
| [110] | RNN | Predicting procedure in business |

Fombellida et al [108] developed a work engaged the concept of meta plasticity, which has the capability of improving the flexibility of learning mechanism, to detect a deeper useful information and learning from data. The study focused on MLP where the output is in the application of BI while utilizing the client data. The developed work approved that the model reinforces learning over the classical MLPs and other systems in terms of learning evolution and ultimate accomplishment regarding to precision and stability. Results of the proposed method were compared with MOE developed by West [111] and MCQP developed by Peng et al [111] in the term of accuracy and sensitivity. Results were presented in Fig. 20.

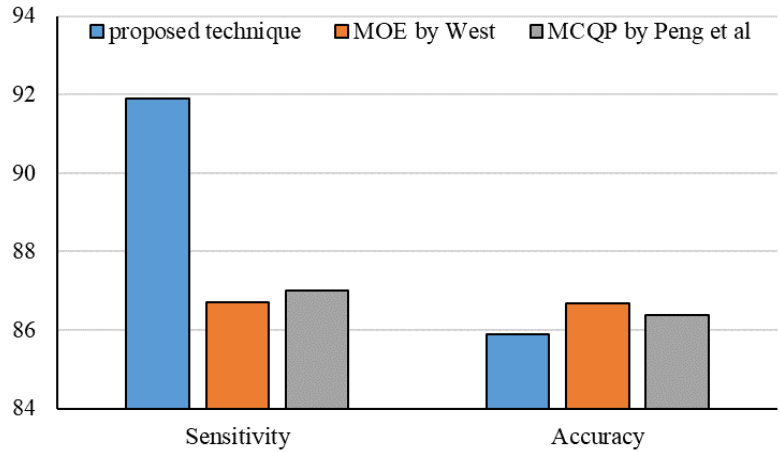

Fig. 20. The visualized results related to the study by Fombellida et al [108]

As is clear from Fig. 20, the proposed method provided a higher sensitivity compared with other methods and lower Accuracy value.

Business intelligence Net, which is a recurrent neural network for exploiting irregularity in business procedure, is designed to mainly manage the data aspect of business procedure. Where this Net does not depend on any given information about the procedure and clean dataset by Nolle et al [109]. The study presented a heuristic setting that decreasing handy workload. The proposed approach can be applied to model the time dimension in sequential phenomenon in which is useful for anomalies purposes. It is proven by the experimental results that BI Net is a trustable approach with high capability of anomalies detection in business phenomenon logs. An alternative strategy to ML algorithm needs to manage huge amounts of data that generated by quicker enlargement and broader use of digital technology. DL enables us to dramatically handle large dataset with heterogeneous property. Singh and Verma [31] designed a new multi-layer feature selection interacting with stacked auto-encoder (SAE) to just detect crucial representations of data. The novel presented method performed better than commonly used ML algorithms applied to the Farm Ads dataset. Results were evaluated by employing accuracy and area under the curve factors. Fig. 21 presents the visualized results.

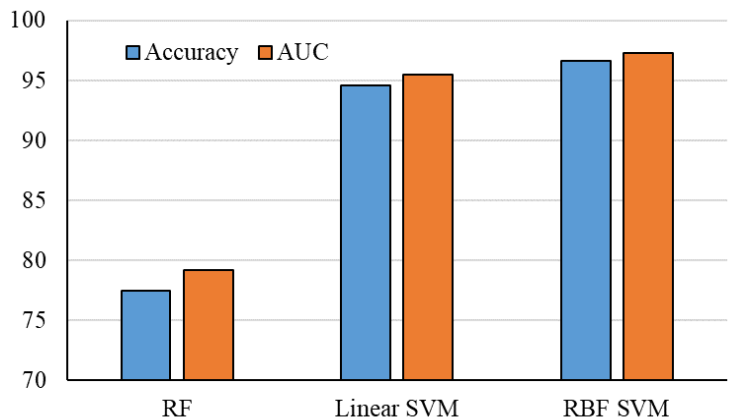

Fig. 21. Results for the study by Singh and Verma [31]

According to Fig. 21, RBF based SVM provided higher accuracy and AUC and the lowest accuracy and AUC was related to RF method. Comparing results for linear SVM and RBF SVM claim that the kernel type is the most important factor for the performance of SVM and RBF kernel provides higher performance compared with that for the linear kernel.

A new approach [110] used recurrent neural networks architecture for forecasting in the business procedure manner. Their approach equipped with explicit process characteristic, no need of model, where the RNN inputs build by means of an embedding space with demonstrated results regarding to the validation accuracy and the feasibility of this method.

The work in [31] utilized novel multi-layer SAE architecture to handle Farm Ads data setting where traditional ML algorithm does not perform well regarding to high dimensionality and huge sample size characteristics of Farm Ads data. Their feature selection approach has the great capability of dimensionality reduction and enhancing the classifier performance. We depicted the complete process of the algorithm in Fig. 22.

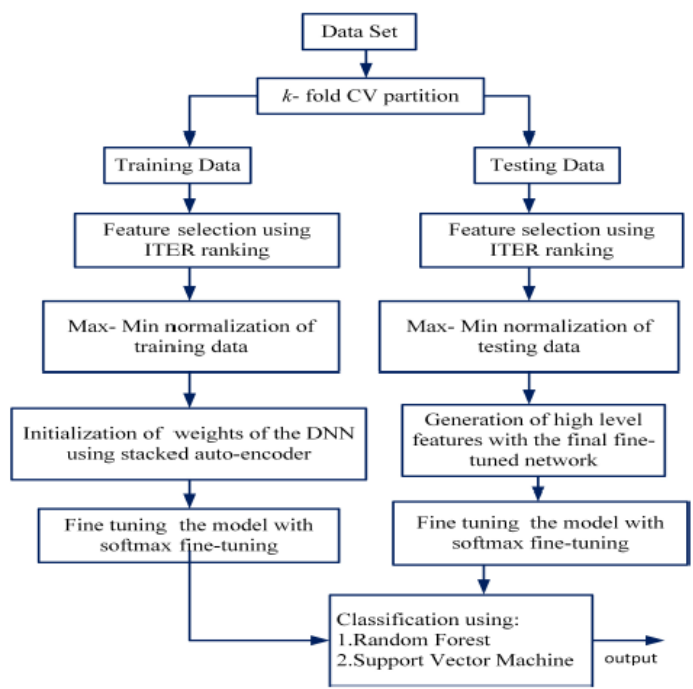

**Fig. 22**. Complete procedure of the proposed model [31]

Although, DL utilize the representation learning property of DNN and handling complex problems involving nonlinear patterns of economic data. But namely, DL methods generally suffer to efficiently align the learning problem with the goal of the trader and to involve significant market constraint to detect the optimal strategy associated with the economics problems. RL approaches enable us to fill the gap with its powerful mathematical structure.

*Deep Reinforcement Learning Application in Economics*

Despite the traditional approaches, DRL has the important capability of capturing substantial market conditions to provide the best strategy in economic, which also provides the potential of scalability and efficiently handling high dimensional problems. Thus, we are motivated to consider the recent advance of deep RL application in economics and financial market.

*Deep Reinforcement Learning in Stock Trading*

Financial companies heavily need to detect the optimal strategy while dealing with stock trading in the dynamic and complicated environment in order to maximize their revenue. Traditional methods applied to stock market trading are quite difficult to be experimented when practitioner wants to in case of considering transaction costs. As well as RL approaches are not efficient enough to find the best strategy due to the lack of scalability of the models to handle high-dimensional problems [113]. Table 13 presents the most notable studies developed by deep RL in stock market.

Table 13. Application of deep RL in stock market.

| References | Methods | Application |
|---|---|---|
| [114] | DDPG | Dynamic stock market |
| [115] | Adaptive DDPG | Stock portfolio strategy |
| [116] | DQN methods | Efficient market strategy |
| [117] | RCNN | Automated trading |

Xiong et al [114] used Deep Deterministic Policy Gradient (DDPG) algorithm as an alternative to explore the optimal strategy in dynamic stock market. While the algorithm components handle large action-state space, taking care of the stability, removing samples correlation and enhancing data utilization. Results demonstrated that the applied model is robust in terms of equilibrating risk and perform better as compared to traditional approaches with the guarantee of the higher return. Xinyi et al [115] devoted to design a new Adaptive Deep Deterministic Reinforcement Learning framework (Adaptive DDPG) dealing with detecting optimal strategy in dynamic and complicated stock market. The model combined optimistic and pessimistic Deep RL that relies on the both negative and positive forecasting errors. Under complicated market situations, the model indicated that has the capability to gain better portfolio profit based on the Dow Jones stocks.

Li et al [116] did survey studies in deep RL for analyzing the multiple algorithms for stock decision-making mechanism. Their experimental results, based on three classical models such as DQN, Double DQN and Dueling DQN, showed that among them, DQN models enable us to attain better investment strategy in order to optimize the return in stock trading where applying empirical data to validate the model. Azhikodan et al [117] focused on automating oscillation in securities trading using deep RL where they employed recurrent convolutional neural network (RCNN) approach for forecasting stock value from the economic news. The original goal of the work was to demonstrate that deep RL techniques are able to efficiently detecting the stock trading tricks. The recent deep RL results in the application of stock market can be viewed in Table 13. An adaptive DDPG [115] comprising both actor network $\mu(s|\theta^\mu)$ and critic network $Q(s,a|\theta^Q)$ utilized for the stock portfolio with different update structure than DDPG approach as given by Eq. 30:

$$\theta^{\acute{Q}} \leftarrow \tau\theta^Q + (1-\tau)\theta^{\acute{Q}}. \tag{30}$$

$$\theta^{\acute{\mu}} \leftarrow \tau\theta^\mu + (1-\tau)\theta^{\acute{\mu}} \tag{31}$$

Where the new target function is Eq.31:

$$\mathcal{Y}_i = r_i + \gamma\acute{Q}(s_{i+1}, \acute{\mu}(s_{i+1}|\theta^{\acute{\mu}}, \theta^{\acute{Q}})) \tag{31}$$

Experimental results indicated that the model outperform considerably the DDPG with high portfolio return as compared to the previous methods. One can check the structure of the model in Fig. 23.

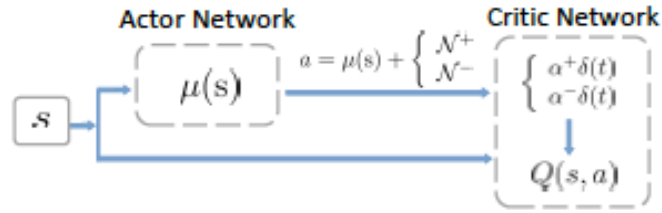

**Fig. 23**. Proposed model architecture [115]

Deep Reinforcement Learning in Portfolio Management

Algorithmic trading area is currently using deep RL techniques to portfolio management while doing the procedure of fixed allocating of a capital into various financial products (see Table 14). Table 4 presents the notable studies in the application of deep reinforcement learning in portfolio management.

Table 14. Application of deep reinforcement learning in portfolio management.

| References | Methods | Application |
| --- | --- | --- |
| [114, 118] | DDPG | Algorithmic trading |
| [119] | Model-less CNN | Financial portfolio algorithm |
| [15] | Model-free | Advanced strategy in portfolio trading |
| [120] | Model-based | Dynamic portfolio optimization |

Liang et al [118] engaged different RL methods like, DDPG, Proximal Policy Optimization (PPO) and PG methods, to obtain the strategy associated with financial portfolio in continuous action-space. They compared the performance of the models in various setting in conjunction with China asset market and indicated that PG is more favorable in stock trading than the two others. The study also presented a novel Adversarial Training approach for ameliorate the training efficiency and the mean-return. Where the new training approach meliorated the performance of PG as compared to Uniform Constant Rebalanced Portfolios (UCRP). A recent work by Jiang et al [119] employed deterministic deep reinforcement learning approach based on a cryptocurrency to obtain optimal strategy in financial problem setting. While Cryptocurrencies, like Bitcoin, can be used in place of governmental money. The study designed model-less convolutional neural network (RNN) where the inputs are historical assets prices from a cryptocurrency exchange with the aim to produce the set of portfolio weights. The proposed models try to optimize the return using network reward function in the reinforcement learning manner. The performance of their presented CNN approach obtained good results as compared to other benchmark algorithms in portfolio management. Another trading algorithm was proposed as model-free Reinforcement Learning framework **[15]** where the reward function was engaged by applying a fully exploiting DPG approach so as to optimize the accumulated return. While the model includes Ensemble of Identical Independent Evaluators (EIIE) topology to incorporate large sets of neural-net in terms of weight-sharing. And a Portfolio-Vector Memory (PVM) in order to prevent the gradient destruction problem. As well as an Online Stochastic Batch Learning (OSBL) scheme to consecutively analyze steady inbound market information, in conjunction with a CNN, a LSTM and a RNN models. The results indicated that the models, while interacting with benchmark models, performed better than previous portfolio management algorithms based on a cryptocurrency market database. A novel model-based deep RL scheme was designed by Yu et al **[120]** in the sense of automated trading to take the action and make the decisions sequentially associated with global goal. The model architecture includes an infused prediction module (IPM), a generative adversarial data augmentation module (DAM) and a behavior cloning module (BCM), dealing with designed back-testing. Empirical results, using historical market data, proved that the model is stable and gain more return as compared to baseline approaches and other recent model-free methods. Portfolio optimization is a challenging task while trading stock in the market. In the Work [120] utilized a novel efficient RL architecture associated with risk sensitive Portfolio. Where combining, IPM to forecast the Stock trend with historical asset prices for improving the performance of RL agent, DAM to control over-fitting problem and BCM to handle unexpected movement in portfolio Weights and to retain the portfolio with low volatility (see Fig. 24). The results demonstrate the complex model is more robust and profitable as compared to the prior approaches.

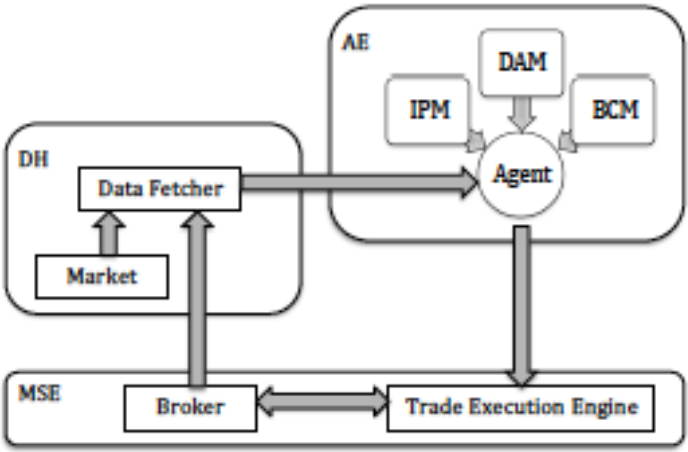

**Fig. 24**. Trading architecture [120]

Deep Reinforcement Learning in Online Services

In current development of online services, the users face to the challenge of detecting their interested items efficiently. Where recommendation techniques enable us to give the right solutions to this problem. Various recommendation methods are presented like content-based collaborative filtering, factorization machines, multi-armed bandits, and to name a few. But considering these proposed approaches are mostly limited to where the users and recommender systems interact statically and focus on short-term reward. Table 15 presents the notable studies in the application of deep reinforcement learning in online services.

Table 15. Application of deep reinforcement learning in online services.

| References | Methods | Application |
| --- | --- | --- |
| [121] | Actor-critic method | Recommendation architecture |

| [122] | SS-RTB method | Bidding optimization in advertising |
| [123] | DDPG and DQN | Pricing algorithm for online market |
| [124] | DQN scheme | Online news recommendation |

Feng et al [121] applied a new recommendation using actor-critic model in RL that can explicitly grab a dynamic interaction and long-term rewards in consecutive decision making process. The experimental work utilizing practical datasets proved that the presented approach outperforms the prior methods. A key task in online advertising is to optimize advertisers gain in the framework of bidding optimization.

Zhao et al [122] focused on real-time bidding (RTB) applied to sponsored search (SS) auction in complicated stochastic environment associated with user action and bidding policies (see Table 15). Their model so-called SS-RTB engaged reinforcement learning concepts to adjust a robust Markov Decision Process (MDP) model in changing environment based on suitable aggregation level of dataset from auction market. Empirical results of both online and offline evaluation based on Alibaba auction platform indicated the usefulness of the proposed method. In the rapid growing of business, online retailers face to the more complicated operations that heavily require to detect updated pricing methodology in order to optimize their profit. Liu et al [123] proposed a pricing algorithm that model the problem in the framework of MDP based on E-commerce platform. They used deep reinforcement learning approaches due to effectively dealing with dynamic market environment and market changes and to set efficient reward function associated with the complex environment. While online testing for such model is impossible as legally the same prices have to be presented to the various customers simultaneously. The model engages deep RL techniques to maximize the long-term return while applying to both discrete and continuous setting for pricing problem. Empirical experiments showed that the obtained policies from DDPG and DQN methods accomplished much better than other pricing strategies based on various business database. Based on reports of Kompan and Bieliková [125] News aggregator services is the favorite online services that able to supply massive volume of content for the users. Therefore, it is important to design news recommendation approaches for boosting user experience. Results were evaluated by precision and recall values into two scenarios including SME.SK and manually annotated. Fig. 25 presents the visualized results for the study by Kompan and Bieliková [125].

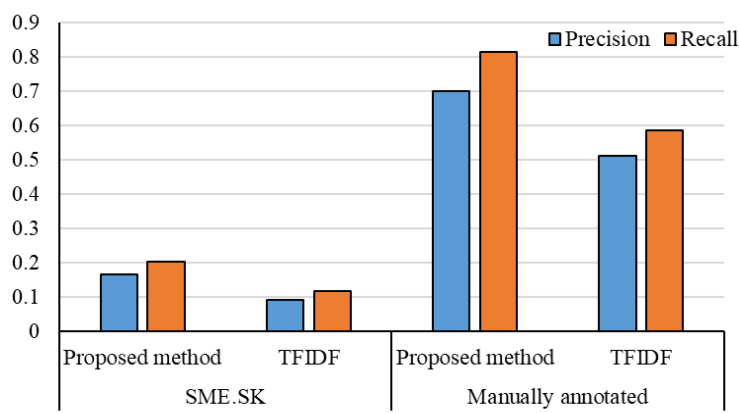

Fig. 25. The visualized results for the study by Kompan and Bieliková [125].

As is clear from Fig. 25, in both scenarios, the proposed method provided the lowest recall and precision compared with TFIDF. A recent research by Zheng et al [124] designed a new framework using DQN scheme for online news recommendation with capability of taking both current and future reward. While the study considered user activeness and also Dueling Bandit Gradient Descent approach to meliorate the recommendation accuracy. Wide empirical results, based on online and offline tests, indicated the superiority of their novel approach.

To deal with the bidding optimization problem is the real-world challenge in online advertising. Despite the prior methods, recent work [118] used an approach called SS-RTB to handle sophisticated changing environment associated with bidding policies (see Table 16). More importantly, the model was extended to the multi-agent problem using robust MDP and then evaluated by the performance metric PUR_AMT/COST that showed the model considerably outperform the PUR_AMT/COST. Where the PUR_AMT/COST is a metric of optimizing buying amount with high correlation of minimizing the cost.

Table 16. Learning algorithm [118].

| Algorithm t DQN Leaming |
|---|
| I: for epsode = 1 to n do |
| 2: Initialize replay memory D to capacity N |
| 3: Initialize action value functions (Qtrain.Qep. Qtarget ) with weights Otrain.Oep.Otarget |
| 4: for t = 1 to m do |
| 5: With probability E select a random action at |
| 6: otherwise select at = arg max 0 Qep(St, a;Θep ) |
| 7: Execute action at to auction simulator and observe state $S_{t+1}$ and reward $r_t$ |
| 8: if budget of $S_{t+1} < 0$, then continue |
| 9: Store transition (s1,at,r1, s1+1 ) in D |
| 10: Sample random rmru batch of transitions (sj,aj,rj, Sj+1 ) from D |
| II: if j = m then |
| 12: Set $Y_i = r_i$ |
| 13: else |
| 14: Set $Y_i = r_i + y$ arg max$_a$, $Q_{target}$ ($s_{i+1}$• a.,$Θ_{target}$ ) |
| 15: end if |
| 16: Perform a gradient descent step on the loss function $(Y_i - Q_{train}( s_i.a_i;Θ_{train}))^2$ |
| 17: end for |
| 18: Update $Θ_{ep}$ withΘ |
| 19: Every C steps, update $Θ_{target}$ with Θ |
| 20: end for |

**Discussions**

A comparative result of DL and DRL models applied to the multiple economic domains discussed in Table 13. Different aspect of proposed models summed up such as model complexity, accuracy and speed, source of dataset, internal detection property of the models, and profitability of the model regarding to revenue and managing the risk. Our work indicated that both DL and DRL algorithms are useful for prediction purpose with almost the same quality of prediction in terms of statistical error. Deep learning models like, AE method in risk management [97] and LSTM-SVR approach [101] in investment problem, showed that they enable agents to considerably maximize their revenue while taking care of risk constraints with reasonable high performance as it is quite important to the economic markets. In addition, reinforcement learning is able to simulate more efficient models with more realistic market constraints. While deep RL goes further that solve the scalability problem of RL algorithms, which is crucial with fast users and market growing, and that efficiently work with high-dimensional settings as it is highly desired in the financial market. Where DRL can give notable helps to design more efficient algorithms for forecasting and analyzing the market with real-word parameters. Deep Deterministic Policy Gradient (DDPG) method used by [114] in stock trading demonstrated that how the model is able to handle large setting concerning stability, improving data using, and equilibrating risk while optimizing the return with high performance guarantee. Another example in deep RL framework made use of DQN scheme [124] for meliorate the news recommendation accuracy dealing with huge users at the same time with considerably high performance guarantee of the method. Our review demonstrated that there is great potential of improving deep learning methods applied to the RL problems to better analyze the relevant problem for finding the best strategy in a wide range of economic domains. Furthermore, RL is highly growing research in DL. However, most of the well-known results in RL have been in single agent environments while the cumulative reward just involves the single state-action spaces. It would be interesting that considering multi-agent scenarios in RL that comprising more than one agent where the cumulative reward can be affected by the actions of other agents. Although, there were some recent research that applied multi-agent reinforcement learning (MARL) scenarios to a small set of agents but not to a large set of agents[126]. For instance, in the financial markets where there are a huge number of agents at the same time, the acts of particular agents, concerning the optimization problem, might be affected by the decisions of the other agents. The MARL can provide an imprecise model and then inexact prediction while considering infinite agents. On the other hand, mathematicians use the theory of mean field games (MFGs) to model a huge variety of non-cooperative agents in a complicated multi-agent dynamic environment. But in theory, MFG modeling might often lead to derive unsolvable partial differential equations (PDE) while dealing with large numbers of agents. Therefore, the big challenge in the application of the MARL is to efficiently model a huge number of agents and then solve the real-world problem using MFGs, dealing with economics and financial markets. Now, the big challenges might be applying MFGs to MARL systems while involving infinite agents. Note that MGFs is able to efficiently model the problem but giving the unsolvable equations [127, 128]. Recent advanced research demonstrated that using MFGs can diminish the complexity of the MARL system dealing with infinite agents. Thus, the question is that how to effectively present the combination of MGFs (mathematically) and MARL that can help out to solve unsolvable equations applied to economic and financial problems. For example, the goal of all the high-frequency traders is to dramatically optimize the profit where MARL scenarios can be modelled by MFGs assuming that all the agents have the same cumulative reward.

Table 13. Comparative study of DL and DRL models in economics

| Methods | Dataset type | Complexity | Detection efficiency | Accuracy & processing rate | Profit & Risk | Application |
|---|---|---|---|---|---|---|
| TGRU | Historical | High | Reasonable | High-Reasonable | Reasonable profit | Stock pricing |
| DNN | Historical | Reasonably high | Reasonable | High-Reasonable | Reasonable profit | Stock pricing |
| AE | Historical | High | High | High-Reasonably high | Reasonably low risk | Insurance |
| RgretNet | Historical | High | Reasonable | High-Reasonable | High profit | Auction design |
| AE & RBM | Historical | Reasonably high | Reasonable | High-Reasonable | Low risk | Credit card fraud |
| ENDE | Historical | Reasonable | Reasonable | High-Reasonable | --- | Macroeconomic |
| AE | Historical | High | High | High-Reasonable | High-Low | Risk management |
| LSTM-SVR | Historical | High | Reasonable | Reasonably-high-High | High-Low | Investment |
| CNNR | Historical | Reasonable | High | Reasonably high-High | Reasonably high | profit Retail market |
| RNN | Historical | Reasonable | High | Reasonable-Reasonable | --- | Business intelligence |
| DDPG | Historical | High | High | Reasonably high-High | High-Low | Stock trading |
| IMF&MBM | Historical | High | High | Reasonabley high-High | High profit | portfolio managing |
| DQN | Historical | High | High | High-High | High-Low | Online services |

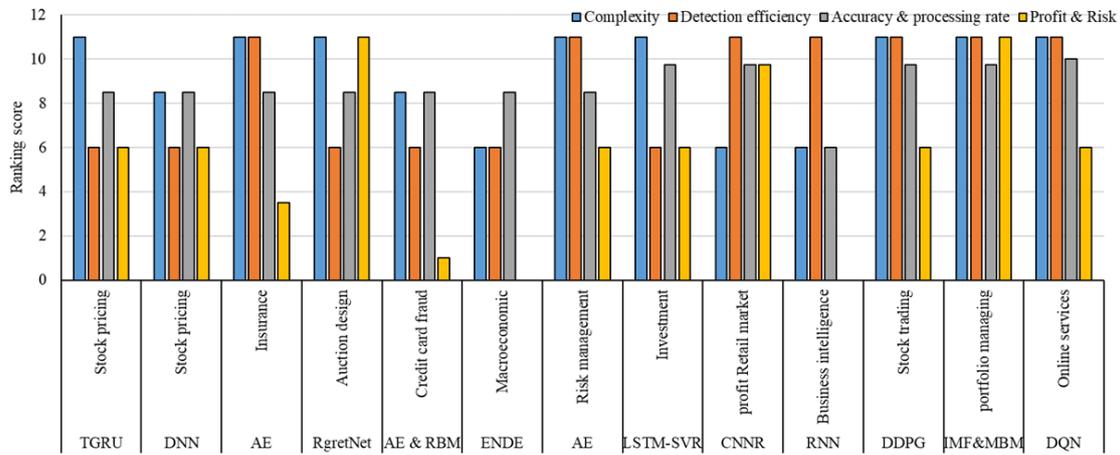

Fig. 26. Ranking score for each method in its own application

**Conclusion**

In the current fast economics and market growth, there is a high demand for the appropriate mechanisms in order to considerably enhance the productivity and quality of the product. Thus, deep learning (DL) can contribute to effectively forecast and detect complex market trend, as compared to the traditional ML algorithms, with the major advantage of high-level features extraction property and proficiency of the problem solver methods. Furthermore, reinforcement learning (RL) enables us to construct more efficient framework regarding to integrate the prediction problem with the portfolio structure task, considering crucial market constraints and better performance. While using deep reinforcement learning architecture, the combination of both DL and RL approaches, for RL to resolve the problem of scalability and to be applied to the high-dimensional problems as desired in real-world market setting. Several DL and deep RL approaches, such as DNN, Autoencoder, RBM, LSTM-SVR, CNN, RNN, DDPG, DQN, and a few others, reviewed in the various application of economic and market domains. Where the advanced models engaged to provide improved prediction, to extract better information and to find optimal strategy mostly in complicated and dynamic market conditions. This brief work represents that the basic issue of the all proposed approaches are mainly to fairly deal with the model complexity, robustness, accuracy, performance, computational tasks, risk constraints and profitability. Practitioners can employ a variety of both DL and deep RL techniques, with the relevant strengths and weaknesses, that serves for the economic problems to enable the machine for detecting the optimal strategy associated with the market. Recent works showed that the novel techniques in DNN, and recently interacting with reinforcement learning so-called deep RL, has the potential to considerably enhance the model performance and accuracy while handling real-world economic problems. We mention that our work indicates the recent approaches in both DL and deep RL perform better than the classical ML approaches. Significant progression can be obtained by designing more efficient novel algorithms using deep neural architectures in conjunction with reinforcement learning concepts to detect the optimal strategy, namely optimize the profit and minimize the loss while concerning the risk parameters, in a highly competitive market.

**Author Contributions:** Equal contribution.

**Funding:** There were no funding for this work.

**Acknowledgments:** We acknowledge the financial support of this work by the Hungarian State and the European Union under the EFOP-3.6.1-16-2016-00010 project and the 2017-1.3.1-VKE-2017-00025 project.

**Conflicts of Interest:** The authors declare no conflict of interest.

**Acronyms**

| | |
|---|---|
| AE | Autoencoder |
| ALT | Augmented Lagrangian Technique |
| ARIMA | Autoregressive Integrated Moving Average |
| BCM | Behavior Cloning Module |
| BI | Business Intelligence |
| CNN | Convolutional Neural Network |
| CNNR | Convolutional Neural Network Regression |
| DAM | Data Augmentation Module |
| DBNs | Deep Belief Networks |
| DDPG | Deterministic Policy Gradient |
| DDQN | Double Deep Q-network |
| DDRL | Deep Deterministic Reinforcement Learning |
| DL | Deep Learning |
| DNN | Deep Neural Network |
| DPG | Deterministic Policy Gradient |
| DQN | Deep Q-network |
| DRL | Deep Reinforcement Learning |
| EIIE | Ensemble of Identical Independent Evaluators |
| ENDE | Encoder-Decoder |
| GANs | Generative Adversarial Nets |
| IPM | Infused Prediction Module |
| LDA | Latent Dirichlet Allocation |
| MARL | Multi-agent Reinforcement Learning |
| MCTS | Monte-Carlo Tree Search |
| MDP | Markov Decision Process |
| MFGs | Mean Field Games |
| ML | Machine Learning |
| MLP | Multilayer Perceptron |
| MLS | Multi Layer Selection |
| NFQCA | Neural Fitted Q Iteration with Continuous Actions |
| NLP | Natural Language Processing |
| OSBL | Online Stochastic Batch Learning |
| PCA | Principal Component Analysis |
| PDE | Partial Differential Equations |
| PG | Policy Gradient |
| PPO | Proximal Policy Optimization |
| PVM | Portfolio-Vector Memory |
| RBM | Restricted Boltzmann Machine |
| RCNN | Recurrent Convolutional Neural Network |
| RL | Reinforcement Learning |
| RNN | Recurrent Neural Network |
| R-NN | Random Neural Network |
| RS | Risk Score |
| RTB | Real-Time Bidding |
| SAE | Stacked Auto-encoder |
| SPF | Survey of Professional Forecasters |
| SPG | Stochastic Policy Gradient |
| SS | Sponsored Search |
| SVR | Support Vector Regression |
| TGRU | Two-stream Gated Recurrent Unit |
| UCRP | Uniform Constant Rebalanced Portfolios |